# Long-term Study of Changes in the Orbital Periods of 18 Eclipsing SW Sextantis Stars

**David Boyd**
*West Challow Observatory, OX12 9TX, UK; davidboyd@orion.me.uk*

**Abstract** SW Sex stars are an informal sub-class of eclipsing nova-like cataclysmic variables. We report 934 new eclipse times measured over the past 17 years for HS 0728+6738 (V482 Cam), SW Sex, DW UMa, HS 0129+2933 (TT Tri), V1315 Aql, PX And, HS 0455+8315, HS 0220+0603, BP Lyn, BH Lyn, LX Ser, UU Aqr, V1776 Cyg, RW Tri, 1RXS J064434.5+334451, AC Cnc, V363 Aur, and BT Mon. When combined with published eclipse times going back in some cases many decades, we show that these binary systems exhibit a range of behaviors, including increasing, decreasing, and possibly oscillating orbital periods. Nevertheless, the duration of these observations is still not long enough to be able to make reliable quantitative statements about their long term behaviors. In addition to these long term trends, we also observed rapid and unusual decreases in the orbital periods of SW Sex and RW Tri during 2017 and 2018, respectively.

## 1. The SW Sex phenomenon

Nova-like variables are a sub-category of cataclysmic variables (CVs) in which the transfer of hydrogen-rich material from the main sequence secondary star to the white dwarf primary via Roche lobe overflow is sustained at a high rate. This maintains the accretion disc around the primary in a bright state and inhibits the disc instability mechanism responsible for dwarf nova outbursts. The majority of nova-like variables have binary orbital periods longer than 3 hours, which places them above the period gap and in the regime where magnetic braking progressively shrinks the binary orbit and drives mass transfer. Further information on CVs can be found in Patterson (1984), Warner (1995), and Hellier (2001).

The name SW Sex stars was first introduced in Thorstensen *et al.* (1991) to characterise a range of observational properties shared by a number of eclipsing nova-like variables which displayed complex and unusual spectral variation with orbital phase. Prototypes of this informal sub-class were SW Sex, DW UMa, PX And, and V1315 Aql. Honeycutt *et al.* (1986) first noticed that SW Sex (known at the time as PG 1012-029) showed deep eclipses in its continuum but hardly at all in its emission lines, suggesting the presence of a bipolar wind emanating from the accretion disk. Several more so-called SW Sex stars were first identified as variables in the Hamburg Quasar Survey (Hagen *et al.* 1995). The observational characteristics of SW Sex stars are described in Hoard *et al.* (2003). Although initially quite narrow, the definition of SW Sex stars now encompasses most nova-like CVs above the period gap with high mass transfer rates. For a review of our knowledge of the SW Sex phenomenon see Schmidtobreick (2015) and references therein.

SW Sex stars with high orbital inclinations experience deep eclipses which provide a means to measure and monitor their orbital periods. Two motivations for this study, which began in 2006, were to produce accurate eclipse ephemerides for predicting future eclipse times and to investigate if any of the stars deviated from the linear ephemeris expected for a constant orbital period. Several of these stars had not been observed systematically for many years and by combining published data on eclipse times going back in some cases over many decades with new eclipse measurements, their ephemerides could be updated and the stability of their orbital periods investigated.

We chose 18 SW Sex stars which are deeply eclipsing, observable from the UK, and bright enough to yield accurate eclipse times with amateur-sized telescopes. These are listed in Table 1 with their mean orbital periods and the time span of available observations including new results reported here. All have orbital periods above the period gap. One member of the group, BT Mon, experienced a nova outburst in 1939 and a nova shell has since been observed (Duerbeck 1987). Nova shells have also been imaged around V1315 Aql (Sahman *et al.* 2015) and AC Cnc (Shara *et al.* 2012), evidence of nova eruptions several hundred years ago. AC Cnc and BT Mon have two of the longest orbital periods in the group.

An initial report covering the period 2006 to 2012 was published in the *Journal of the AAVSO* (Boyd 2012), hereafter referred to as Paper 1. Here we report on a continuation of this study to 2023 and present results which now cover a 17-year period.

## 2. Measuring new eclipse times

Predicted times of primary eclipses were obtained from the ephemerides in Paper 1 and a time-series of images of the field of each star obtained starting well before and ending well after these predicted eclipse times to allow for possible variation in orbital period. All images were made unfiltered to maximize photon statistics with either a 0.25-m or 0.35-m Schmidt-Cassegrain Telescope (SCT) and an SXV-H9 (later SXVR-H9) CCD camera located at West Challow Observatory near Oxford, UK. Image scales with these telescopes were 1.45 and 1.21 arcsec/pixel, respectively. Images were dark subtracted and flat fielded and a magnitude for the star was measured in each image using differential aperture photometry with respect to an ensemble of between three and five nearby comparison stars. Comparison star V band magnitudes with errors were obtained from AAVSO charts or from catalogues available at the start of the study. The same comparison star magnitudes and analysis procedures have been used for each star throughout the study to maintain consistency. A list of comparison stars used for each variable is given in Table 2. If we were starting the project today, we would choose comparison stars from the AAVSO Photometric All-Sky Survey (Henden *et al.* 2018). The photometry error for each star was calculated using the CCD Equation (Howell 2006). For each comparison star this error was then added in quadrature with the comparison chart magnitude error and a weighted mean magnitude zero point and error was computed for the image. This was then used to compute the magnitude and error of the variable star for that image.

A quadratic polynomial was fitted to the lower section of each eclipse in order to find the time of minimum which was expressed as a Heliocentric Julian Date (HJD). An associated analytical error in the time of minimum was derived from uncertainties in the magnitude measurements. The section of the eclipse used for the polynomial fit was normally between the points of maximum slope of the eclipse ingress and egress. Figure 1 shows examples of eclipse profiles. Uncertainties in individual magnitude measurements are generally smaller than the plotted mark. Some eclipses have rounded minima, some are V-shaped, while others exhibit random fluctuations in light output throughout the eclipse, indicating that the source of these fluctuations has not been eclipsed. Irregular eclipse profiles are more difficult to measure and this can lead to larger uncertainties in measured times of minimum. In what follows we will refer to these uncertainties as errors.

It was generally found that analytical errors from the quadratic fits underestimated the real uncertainty in eclipse times. The scatter in eclipse times for each star over a short interval during which the eclipse times were likely to have varied linearly was examined and the analytical errors scaled to make them consistent with the observed scatter about the linear trend. For stars with the smoothest eclipses, a scaling factor of 3 gave errors consistent with the scatter of eclipse times, while for eclipses with the largest fluctuations a factor of 7 was required. This scaling factor was generally found to be consistent for each star throughout the study.

A total of 898 new eclipse times for the 18 stars in this study have been observed and measured by the author. The number of new eclipse times for each star are listed in Table 1. Based on the ephemerides in Paper 1, cycle numbers were assigned to each new eclipse. Measured eclipse times with errors and corresponding cycle numbers for each of the 18 stars are listed in Tables 3.1 to 3.18. For completeness we also include here the eclipse times given in Paper 1. A further 36 eclipse times for LX Ser were measured by the author from observations of LX Ser by Cook and Dvorak in the AAVSO International Database (Kafka 2021). These are listed in Table 4.

## 3. Published eclipse times

Altogether 1338 eclipse times for these 18 stars were found in more than 40 published papers and in many issues of *Information Bulletin on Variable Stars* (IBVS), *Bulletin of the Variable Star Observers League in Japan* (BVSOLJ), and *Open European Journal on Variable Stars* (OEJV). The numbers of published eclipse times for each star are listed in Table 1 and the sources of published eclipse times are given in Table 5. We have not included these already published times here for reasons of space. All times of minimum were expressed in HJD for consistency, including some times originally reported in Barycentric Julian Date (BJD). In several cases errors for these eclipse times were not specified in the literature or the errors given were clearly unrealistically small given the observed spread in eclipse times. In these cases we needed to make a realistic estimate of the error in these eclipse times so they could be included in our analysis with appropriate weights. Each such data set was considered separately and the root-mean-square (rms) residual of all the times in that set calculated with respect to a locally fitted linear ephemeris. This value was then assigned as an error to all the eclipse times in that set.

We found that eclipse times derived from photographic plates generally had a large scatter compared to electronically measured times and in practice did not provide a constraint on fitting an ephemeris, so we decided not to include these in this analysis. Eclipse times for RW Tri in Smak (1995) appeared very discrepant with other times reported around the same period and therefore have not been included in this analysis.

## 4. O–C analysis

Each observed eclipse time of minimum was given a weight equal to the inverse square of its assigned error. A weighted linear fit of all available eclipse times vs cycle numbers was calculated for each star. This linear ephemeris was used to produce a calculated time for each eclipse. The linear term in the ephemeris is the mean binary orbital period of the star over the time interval spanned by the observations. Observed minus calculated (O–C) times for each eclipse were then plotted vs cycle number to produce an O–C diagram for each star.

An apparently linear trend in an O–C diagram is consistent with a constant orbital period, while O–C trajectories curving upward indicate the orbital period is increasing and curving downward that the orbital period is decreasing. In most cases we also calculated a weighted quadratic fit to the O–C values. This quadratic ephemeris gave a mean rate of change of orbital period. In some cases, there was a suggestion of sinusoidal variation relative to a linear ephemeris or quadratic ephemeris. In these cases, a weighted sinusoidal fit was calculated with respect to the linear or quadratic ephemeris.

Table 6 gives weighted linear ephemerides for each star computed as described above. SW Sex experienced a large change in its behavior in 2017 and two linear ephemerides are given for before and after this change. Table 7 gives weighted quadratic ephemerides and mean rates of period change for stars where these were calculated.

Our effort to make the weights used in these fits more realistic will inevitably have introduced an element of subjectivity. Therefore we do not compute a quantitative goodness of fit metric such as a reduced chi-squared for each fit as this would not be an objective basis for evaluating fit quality. This is particularly true in the case of a nonlinear model where there are recognized problems in interpreting such a metric (Andrae *et al.* 2010).

Table 1. Eclipsing SW Sex stars in this study.

| *Star name* | $P_{orb}$ *(hours)* | *Time span of obs. (years)* | *New eclipse times measured in this study* | *Previously published eclipse times* |
|---|---|---|---|---|
| HS 0728+6738 = V482 Cam | 3.21 | 20 | 44 | 13 |
| SW Sex = PG 1012-029 | 3.24 | 43 | 49 | 131 |
| DW UMa = PG1030+590 | 3.28 | 39 | 58 | 596 |
| HS 0129+2933 = TT Tri | 3.35 | 20 | 42 | 30 |
| V1315 Aql | 3.35 | 38 | 51 | 80 |
| PX And = PG0027+260 | 3.51 | 31 | 45 | 44 |
| HS 0455+8315 | 3.57 | 21 | 44 | 9 |
| HS 0220+0603 | 3.58 | 20 | 37 | 13 |
| BP Lyn = PG0859+415 | 3.67 | 32 | 45 | 16 |
| BH Lyn = PG0818+513 | 3.74 | 31 | 43 | 33 |
| LX Ser = Stepanyan's Star | 3.80 | 42 | 82 * | 74 |
| UU Aqr | 3.93 | 37 | 53 | 53 |
| V1776 Cyg = Lanning 90 | 3.95 | 35 | 58 | 11 |
| RW Tri | 5.57 | 65 | 58 | 151 |
| 1RXS J064434.5+334451 | 6.47 | 18 | 70 | 36 |
| AC Cnc | 7.21 | 41 | 49 | 19 |
| V363 Aur = Lanning 10 | 7.71 | 42 | 62 | 19 |
| BT Mon | 8.01 | 45 | 44 | 10 |
| Total | | 616 | 934 | 1338 |

*Note: * Includes 36 eclipse times for LX Ser measured by the author from observations of LX Ser by Cook and Dvorak in the AAVSO International Database.*

Table 2. Comparison stars used to measure the time of minimum for each star.

| *Star Name* | *Comparison Stars Used* |
|---|---|
| HS 0728+6738 = V482 Cam | GSC 4360 0033, GSC 4124 0603 |
| SW Sex = PG 1012-029 | GSC 4907 1166, GSC 4907 0207, 2MASS J10145841-0305432 |
| DW UMa = PG1030+590 | GSC 3822 0070, GSC 3822 0983, GSC 3822 1157 |
| HS 0129+2933 = TT Tri | GSC 1755 0855, GSC 1755 0871, GSC 1755 0942, GSC 1755 0926, GSC 1755 0982 |
| V1315 Aql | GSC 1049 1329, GSC 1049 1288, GSC 1049 0464 |
| PX And = PG0027+260 | GSC 1734 0906, GSC 1734 1620, GSC 1734 0752 |
| HS 0455+8315 | GSC 4617 1102, GSC 4617 0542, 2MASS J05071087+8318101, 2MASS J05084059+8316305, 2MASS J 05041189+8321282 |
| HS 0220+0603 | GSC 0045 1418, GSC 0045 0338, GSC 0045 1226, GSC 0045 1400, GSC 0045 0626 |
| BP Lyn = PG0859+415 | GSC 2986 1255, GSC 2986 1258, GSC 2986 1413, GSC 2986 1427 |
| BH Lyn = PG0818+513 | GSC 3421 1055, GSC 3421 0865, GSC 3421 1015 |
| LX Ser = Stepanyan's Star | GSC 1497 1576, GSC 1497 0962, GSC 1497 1643, [HH95] LX Ser-4, [HH95] LX Ser-8 |
| UU Aqr | TYC 5227 0328, GSC 5227 0662, GSC 5227 0399, GSC 5227 0982 |
| V1776 Cyg = Lanning 90 | GSC 3572 1508, 2MASS J20234934+4629294, 2MASS J20234988+4632359, 2MASS J20231931+4629502, 2MASS J20233377+4634165 |
| RW Tri | GSC 1774 0082, GSC 1178 0469, GSC 1774 0357, GSC 1774 0002 |
| 1RXS J064434.5+334451 | [SGH2007] J0644-R, [SGH2007] J0644-S, [SGH2007] J0644-E, [SGH2007] J0644-G, [SGH2007] J0644-M |
| AC Cnc | GSC 0816 1525, GSC 0816 1021, GSC 0816 1547, GSC 0816 0998, GSC 0816 0862 |
| V363 Aur = Lanning 10 | [HH95] V363 Aur-04, [HH95] V363 Aur-19, [HH95] V363 Aur-03 |
| BT Mon | GSC 4803 0262, 2MASS J06433904-0204189, 2MASS J06435331-0202124, 2MASS J06433839-0203003 |

*Note: [HH95] = Henden and Honeycutt (1995), [SGH2007] = Sing et al. (2007)*

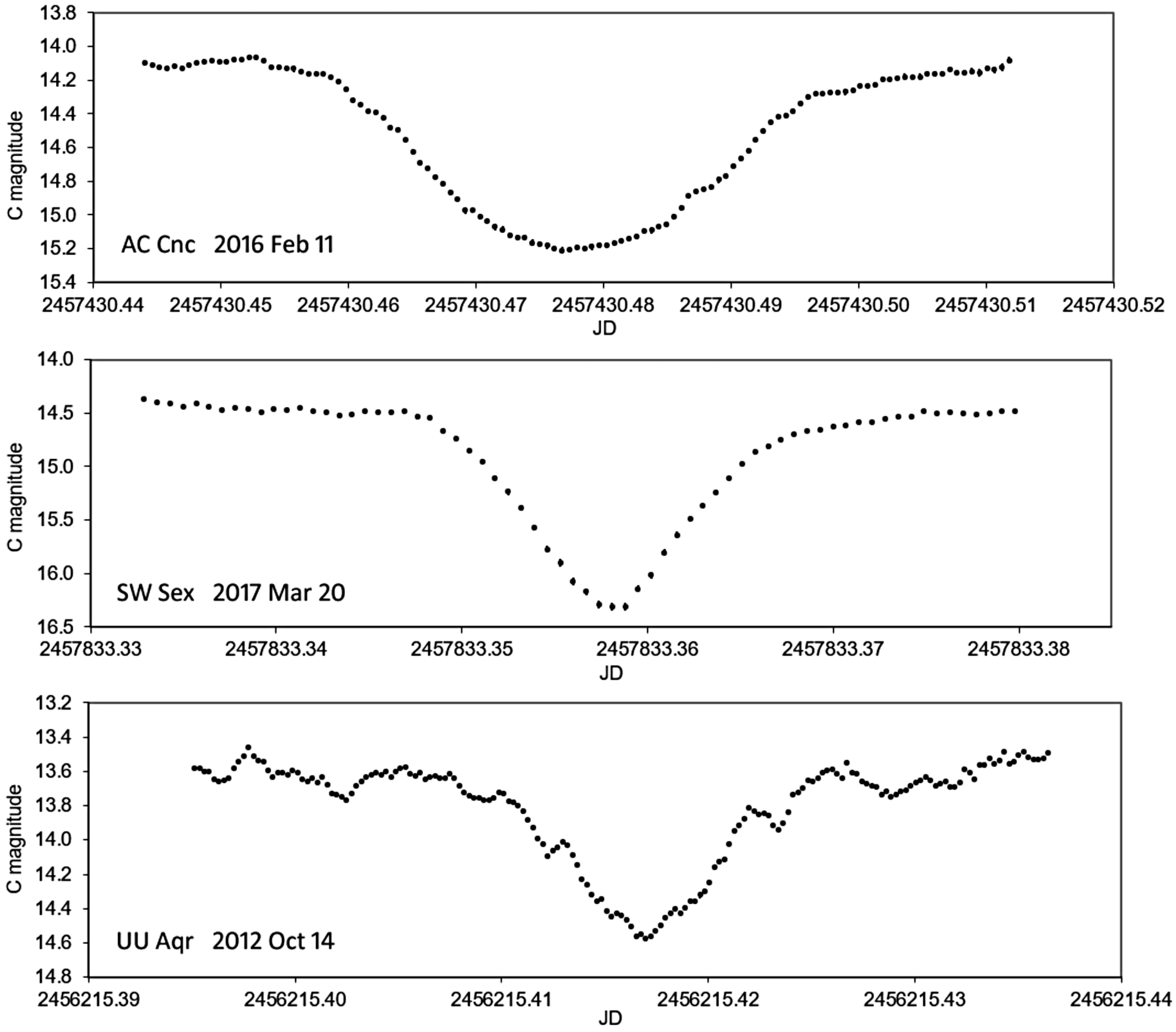

Figure 1. Examples of eclipse profiles. Uncertainties in individual magnitude measurements are generally smaller than the plotted mark.

Table 3.1. Eclipse times, errors and cycle numbers for HS 0728+6738 observed and measured by the author in this study.

| *Eclipse time (HJD)* | *Error (d)* | *Cycle Number* |
|---|---|---|
| 2453810.40077 | 0.00041 | 13539 |
| 2453836.45653 | 0.00024 | 13734 |
| 2453851.42254 | 0.00023 | 13846 |
| 2453853.42648 | 0.00013 | 13861 |
| 2454174.51418 | 0.00022 | 16264 |
| 2454181.32859 | 0.00025 | 16315 |
| 2454185.33706 | 0.00025 | 16345 |
| 2454186.40643 | 0.00024 | 16353 |
| 2454473.42029 | 0.00023 | 18501 |
| 2454493.33001 | 0.00023 | 18650 |
| 2454507.35967 | 0.00039 | 18755 |
| 2454835.39541 | 0.00032 | 21210 |
| 2454891.38182 | 0.00010 | 21629 |
| 2454895.39084 | 0.00009 | 21659 |
| 2454907.41644 | 0.00022 | 21749 |
| 2455188.41832 | 0.00021 | 23852 |
| 2455191.35834 | 0.00014 | 23874 |
| 2455200.31029 | 0.00019 | 23941 |
| 2455515.38459 | 0.00024 | 26299 |
| 2455520.32865 | 0.00038 | 26336 |
| 2455533.42346 | 0.00028 | 26434 |
| 2455889.38551 | 0.00036 | 29098 |
| 2455891.39036 | 0.00024 | 29113 |
| 2455893.39432 | 0.00019 | 29128 |
| 2456267.39501 | 0.00019 | 31927 |
| 2456271.40415 | 0.00015 | 31957 |
| 2456298.39531 | 0.00018 | 32159 |
| 2456725.30918 | 0.00035 | 35354 |
| 2457017.40117 | 0.00013 | 37540 |
| 2457020.34099 | 0.00034 | 37562 |
| 2457442.31106 | 0.00011 | 40720 |
| 2457443.38007 | 0.00017 | 40728 |
| 2458099.31749 | 0.00018 | 45637 |
| 2458103.45938 | 0.00016 | 45668 |
| 2458106.26552 | 0.00040 | 45689 |
| 2458444.32328 | 0.00034 | 48219 |
| 2458477.32701 | 0.00022 | 48466 |
| 2458493.36113 | 0.00005 | 48586 |
| 2458784.38450 | 0.00022 | 50764 |
| 2458806.29770 | 0.00031 | 50928 |
| 2458817.25411 | 0.00029 | 51010 |
| 2459149.43247 | 0.00024 | 53496 |
| 2459157.44964 | 0.00016 | 53556 |
| 2459159.32000 | 0.00023 | 53570 |

Table 3.2. Eclipse times, errors and cycle numbers for SW Sex observed and measured by the author in this study.

| *Eclipse time (HJD)* | *Error (d)* | *Cycle Number* |
|---|---|---|
| 2454185.43702 | 0.00044 | 72965 |
| 2454186.38145 | 0.00029 | 72972 |
| 2454553.41407 | 0.00048 | 75692 |
| 2454564.34410 | 0.00020 | 75773 |
| 2454906.41325 | 0.00019 | 78308 |
| 2454907.49269 | 0.00019 | 78316 |
| 2455260.35696 | 0.00018 | 80931 |
| 2455278.43821 | 0.00012 | 81065 |
| 2455630.35814 | 0.00026 | 83673 |
| 2455660.44910 | 0.00014 | 83896 |
| 2455662.33853 | 0.00028 | 83910 |
| 2455992.39775 | 0.00022 | 86356 |
| 2456005.48662 | 0.00010 | 86453 |
| 2456008.45550 | 0.00018 | 86475 |
| 2456343.50779 | 0.00012 | 88958 |
| 2456354.43764 | 0.00013 | 89039 |
| 2456356.46180 | 0.00016 | 89054 |
| 2456728.35219 | 0.00019 | 91810 |
| 2456739.41702 | 0.00013 | 91892 |
| 2457118.45908 | 0.00019 | 94701 |
| 2457119.40351 | 0.00010 | 94708 |
| 2457461.33764 | 0.00017 | 97242 |
| 2457462.41694 | 0.00015 | 97250 |
| 2457465.38568 | 0.00024 | 97272 |
| 2457833.36314 | 0.00014 | 99999 |
| 2457835.38696 | 0.00014 | 100014 |
| 2457836.33134 | 0.00026 | 100021 |
| 2457837.41089 | 0.00015 | 100029 |
| 2457862.37464 | 0.00018 | 100214 |
| 2458191.48919 | 0.00009 | 102653 |
| 2458212.40447 | 0.00013 | 102808 |
| 2458214.42856 | 0.00010 | 102823 |
| 2458567.42678 | 0.00008 | 105439 |
| 2458571.33929 | 0.00016 | 105468 |
| 2458575.38787 | 0.00011 | 105498 |
| 2458584.42857 | 0.00009 | 105565 |
| 2458585.37339 | 0.00014 | 105572 |
| 2458931.35406 | 0.00015 | 108136 |
| 2458932.43393 | 0.00010 | 108144 |
| 2458933.37801 | 0.00029 | 108151 |
| 2459281.38341 | 0.00015 | 110730 |
| 2459282.46286 | 0.00017 | 110738 |
| 2459291.36857 | 0.00011 | 110804 |
| 2459677.42673 | 0.00014 | 113665 |
| 2459683.36420 | 0.00025 | 113709 |
| 2459685.38844 | 0.00021 | 113724 |
| 2460052.42102 | 0.00008 | 116444 |
| 2460054.44524 | 0.00011 | 116459 |
| 2460064.43065 | 0.00028 | 116533 |

Table 3.3. Eclipse times, errors and cycle numbers for DW UMa observed and measured by the author in this study.

| *Eclipse time (HJD)* | *Error (d)* | *Cycle Number* |
|---|---|---|
| 2454181.41978 | 0.00019 | 58214 |
| 2454185.38111 | 0.00029 | 58243 |
| 2454224.45051 | 0.00043 | 58529 |
| 2454473.34780 | 0.00038 | 60351 |
| 2454564.46466 | 0.00020 | 61018 |
| 2454580.44785 | 0.00033 | 61135 |
| 2454580.58433 | 0.00026 | 61136 |
| 2454588.37104 | 0.00028 | 61193 |
| 2454588.50711 | 0.00019 | 61194 |
| 2454593.42488 | 0.00022 | 61230 |
| 2454596.43092 | 0.00033 | 61252 |
| 2454884.39723 | 0.00024 | 63360 |
| 2454892.32009 | 0.00025 | 63418 |
| 2455239.30026 | 0.00021 | 65958 |
| 2455263.34322 | 0.00014 | 66134 |
| 2455270.31000 | 0.00013 | 66185 |
| 2455278.37037 | 0.00017 | 66244 |
| 2455627.39978 | 0.00017 | 68799 |
| 2455628.35604 | 0.00020 | 68806 |
| 2455629.31205 | 0.00029 | 68813 |
| 2455991.45632 | 0.00028 | 71464 |
| 2456029.43254 | 0.00029 | 71742 |
| 2456033.39472 | 0.00022 | 71771 |
| 2456088.44663 | 0.00035 | 72174 |
| 2456382.42440 | 0.00027 | 74326 |
| 2456384.47293 | 0.00013 | 74341 |
| 2456399.36316 | 0.00039 | 74450 |
| 2456413.43361 | 0.00024 | 74553 |
| 2456728.44826 | 0.00015 | 76859 |
| 2456739.37663 | 0.00011 | 76939 |
| 2457020.37615 | 0.00026 | 78996 |
| 2457021.46907 | 0.00019 | 79004 |
| 2457075.42859 | 0.00021 | 79399 |
| 2457106.43843 | 0.00010 | 79626 |
| 2457108.35065 | 0.00012 | 79640 |
| 2457108.48716 | 0.00020 | 79641 |
| 2458174.42888 | 0.00017 | 87444 |
| 2458188.36286 | 0.00021 | 87546 |
| 2458191.36841 | 0.00019 | 87568 |
| 2458227.43235 | 0.00020 | 87832 |
| 2458231.39341 | 0.00016 | 87861 |
| 2458234.39877 | 0.00035 | 87883 |
| 2458539.44267 | 0.00039 | 90116 |
| 2458540.39817 | 0.00023 | 90123 |
| 2458541.35463 | 0.00015 | 90130 |
| 2458571.40737 | 0.00012 | 90350 |
| 2458585.34131 | 0.00025 | 90452 |
| 2458593.40129 | 0.00019 | 90511 |
| 2458855.41318 | 0.00034 | 92429 |
| 2458861.42412 | 0.00031 | 92473 |
| 2458868.39096 | 0.00013 | 92524 |
| 2458948.44193 | 0.00028 | 93110 |
| 2459258.40274 | 0.00058 | 95379 |
| 2459268.37535 | 0.00016 | 95452 |
| 2459272.33702 | 0.00017 | 95481 |
| 2459597.32418 | 0.00032 | 97860 |
| 2459599.37245 | 0.00018 | 97875 |
| 2459600.32970 | 0.00029 | 97882 |
| 2459968.34763 | 0.00022 | 100576 |
| 2459975.31458 | 0.00022 | 100627 |
| 2459989.38528 | 0.00017 | 100730 |

*Note: The Tables 3.1 through 3.18 are available through the AAVSO ftp site at ftp://ftp.aavso.org/public/datasets/3882-Boyd-511-swsex.txt (if necessary, copy and paste link into the address bar of a web browser).*

Table 3.4. Eclipse times, errors and cycle numbers for HS0129+2933 observed and measured by the author in this study.

| *Eclipse time (HJD)* | *Error (d)* | *Cycle Number* |
|---|---|---|
| 2454061.46332 | 0.00014 | 10892 |
| 2454081.29219 | 0.00016 | 11034 |
| 2454086.45848 | 0.00008 | 11071 |
| 2455106.37036 | 0.00038 | 18375 |
| 2455188.47729 | 0.00030 | 18963 |
| 2455191.27007 | 0.00019 | 18983 |
| 2455460.49099 | 0.00013 | 20911 |
| 2455533.38206 | 0.00022 | 21433 |
| 2455827.45860 | 0.00014 | 23539 |
| 2455835.41776 | 0.00016 | 23596 |
| 2455836.39518 | 0.00010 | 23603 |
| 2456200.43010 | 0.00037 | 26210 |
| 2456215.37178 | 0.00019 | 26317 |
| 2456237.29459 | 0.00022 | 26474 |
| 2456527.46137 | 0.00028 | 28552 |
| 2456611.38335 | 0.00017 | 29153 |
| 2456901.41023 | 0.00021 | 31230 |
| 2456904.48240 | 0.00024 | 31252 |
| 2457258.46323 | 0.00012 | 33787 |
| 2457276.47630 | 0.00017 | 33916 |
| 2457624.45255 | 0.00033 | 36408 |
| 2457631.43448 | 0.00018 | 36458 |
| 2458029.40123 | 0.00026 | 39308 |
| 2458054.39650 | 0.00011 | 39487 |
| 2458056.35143 | 0.00013 | 39501 |
| 2458362.43603 | 0.00029 | 41693 |
| 2458363.41333 | 0.00016 | 41700 |
| 2458388.40829 | 0.00022 | 41879 |
| 2458721.44286 | 0.00017 | 44264 |
| 2458741.41060 | 0.00012 | 44407 |
| 2458759.42431 | 0.00004 | 44536 |
| 2458773.38793 | 0.00031 | 44636 |
| 2458906.32283 | 0.00025 | 45588 |
| 2459105.44446 | 0.00018 | 47014 |
| 2459106.42148 | 0.00027 | 47021 |
| 2459107.39970 | 0.00014 | 47028 |
| 2459523.37876 | 0.00015 | 50007 |
| 2459526.31147 | 0.00011 | 50028 |
| 2459541.39281 | 0.00015 | 50136 |
| 2459914.36388 | 0.00021 | 52807 |
| 2459921.34628 | 0.00025 | 52857 |
| 2459928.32801 | 0.00018 | 52907 |

Table 3.5. Eclipse times, errors and cycle numbers for V1315 Aql observed and measured by the author in this study.

| *Eclipse time (HJD)* | *Error (d)* | *Cycle Number* |
|---|---|---|
| 2454272.50437 | 0.00018 | 59916 |
| 2454306.44865 | 0.00027 | 60159 |
| 2454313.43262 | 0.00072 | 60209 |
| 2454651.48330 | 0.00048 | 62629 |
| 2454670.48100 | 0.00046 | 62765 |
| 2454810.31097 | 0.00082 | 63766 |
| 2455004.47952 | 0.00029 | 65156 |
| 2455006.43480 | 0.00049 | 65170 |
| 2455038.42351 | 0.00055 | 65399 |
| 2455052.39293 | 0.00070 | 65499 |
| 2455463.36184 | 0.00047 | 68441 |
| 2455464.33978 | 0.00036 | 68448 |
| 2455490.32143 | 0.00026 | 68634 |
| 2455777.38468 | 0.00040 | 70689 |
| 2455783.39087 | 0.00040 | 70732 |
| 2455903.24546 | 0.00047 | 71590 |
| 2456131.49866 | 0.00042 | 73224 |
| 2456149.51903 | 0.00035 | 73353 |
| 2456150.49660 | 0.00023 | 73360 |
| 2456215.31256 | 0.00061 | 73824 |
| 2456446.49995 | 0.00064 | 75479 |
| 2456453.48465 | 0.00056 | 75529 |
| 2456478.48866 | 0.00025 | 75708 |
| 2456838.47024 | 0.00025 | 78285 |
| 2456845.45485 | 0.00023 | 78335 |
| 2456895.46344 | 0.00033 | 78693 |
| 2457177.49766 | 0.00035 | 80712 |
| 2457184.48150 | 0.00042 | 80762 |
| 2457203.47971 | 0.00026 | 80898 |
| 2457293.30101 | 0.00042 | 81541 |
| 2457303.35804 | 0.00028 | 81613 |
| 2457563.46136 | 0.00028 | 83475 |
| 2457587.48793 | 0.00021 | 83647 |
| 2457590.42138 | 0.00019 | 83668 |
| 2457960.46038 | 0.00063 | 86317 |
| 2457971.49598 | 0.00025 | 86396 |
| 2457978.48056 | 0.00023 | 86446 |
| 2458294.45908 | 0.00057 | 88708 |
| 2458295.43676 | 0.00018 | 88715 |
| 2458314.43516 | 0.00027 | 88851 |
| 2458655.41791 | 0.00044 | 91292 |
| 2458665.47589 | 0.00053 | 91364 |
| 2458666.45373 | 0.00030 | 91371 |
| 2459024.47947 | 0.00032 | 93934 |
| 2459025.45740 | 0.00032 | 93941 |
| 2459033.41968 | 0.00041 | 93998 |
| 2459365.46204 | 0.00035 | 96375 |
| 2459366.44091 | 0.00031 | 96382 |
| 2459379.43190 | 0.00044 | 96475 |
| 2459744.44107 | 0.00062 | 99088 |
| 2459756.45506 | 0.00023 | 99174 |
| 2459757.43293 | 0.00059 | 99181 |

Table 3.6. Eclipse times, errors and cycle numbers for PX And observed and measured by the author in this study.

| *Eclipse time (HJD)* | *Error (d)* | *Cycle Number* |
|---|---|---|
| 2454318.44729 | 0.00051 | 34708 |
| 2454319.47234 | 0.00046 | 34715 |
| 2454325.47261 | 0.00036 | 34756 |
| 2454448.40773 | 0.00061 | 35596 |
| 2454473.28943 | 0.00051 | 35766 |
| 2454503.29163 | 0.00022 | 35971 |
| 2454761.45718 | 0.00049 | 37735 |
| 2454770.38547 | 0.00069 | 37796 |
| 2455064.40680 | 0.00108 | 39805 |
| 2455066.45577 | 0.00069 | 39819 |
| 2455173.29503 | 0.00032 | 40549 |
| 2455186.32065 | 0.00020 | 40638 |
| 2455188.36884 | 0.00125 | 40652 |
| 2455191.29553 | 0.00055 | 40672 |
| 2455201.24653 | 0.00014 | 40740 |
| 2455460.43876 | 0.00028 | 42511 |
| 2455495.26963 | 0.00061 | 42749 |
| 2455515.46733 | 0.00025 | 42887 |
| 2455795.43984 | 0.00024 | 44800 |
| 2455819.44115 | 0.00069 | 44964 |
| 2455823.39248 | 0.00044 | 44991 |
| 2455901.25250 | 0.00064 | 45523 |
| 2456149.46690 | 0.00038 | 47219 |
| 2456159.41895 | 0.00035 | 47287 |
| 2456215.32501 | 0.00053 | 47669 |
| 2456512.42294 | 0.00048 | 49699 |
| 2456518.42177 | 0.00083 | 49740 |
| 2456609.45353 | 0.00040 | 50362 |
| 2456611.35720 | 0.00069 | 50375 |
| 2456908.45223 | 0.00028 | 52405 |
| 2456922.35622 | 0.00047 | 52500 |
| 2457271.40745 | 0.00041 | 54885 |
| 2457275.50498 | 0.00026 | 54913 |
| 2457615.48272 | 0.00042 | 57236 |
| 2457624.41054 | 0.00047 | 57297 |
| 2457994.38914 | 0.00082 | 59825 |
| 2457996.43794 | 0.00026 | 59839 |
| 2457997.46265 | 0.00064 | 59846 |
| 2458362.46761 | 0.00038 | 62340 |
| 2458379.44407 | 0.00025 | 62456 |
| 2458759.37448 | 0.00018 | 65052 |
| 2458806.35537 | 0.00021 | 65373 |
| 2458817.33089 | 0.00013 | 65448 |
| 2459114.42636 | 0.00047 | 67478 |
| 2459148.38126 | 0.00068 | 67710 |

Table 3.7. Eclipse times, errors and cycle numbers for HS 0455+8315 observed and measured by the author in this study.

| *Eclipse time (HJD)* | *Error (d)* | *Cycle Number* |
|---|---|---|
| 2454061.40139 | 0.00016 | 14807 |
| 2454063.48351 | 0.00020 | 14821 |
| 2454078.35643 | 0.00014 | 14921 |
| 2454112.41335 | 0.00017 | 15150 |
| 2454114.49593 | 0.00023 | 15164 |
| 2454115.38831 | 0.00017 | 15170 |
| 2454895.44552 | 0.00018 | 20415 |
| 2454906.45070 | 0.00013 | 20489 |
| 2454907.34318 | 0.00026 | 20495 |
| 2455065.43666 | 0.00029 | 21558 |
| 2455495.39753 | 0.00032 | 24449 |
| 2455519.49112 | 0.00017 | 24611 |
| 2455526.48082 | 0.00018 | 24658 |
| 2455835.38030 | 0.00021 | 26735 |
| 2455850.40114 | 0.00018 | 26836 |
| 2456271.43853 | 0.00015 | 29667 |
| 2456274.41353 | 0.00029 | 29687 |
| 2456294.34258 | 0.00019 | 29821 |
| 2456538.39879 | 0.00012 | 31462 |
| 2456903.36710 | 0.00014 | 33916 |
| 2456908.42377 | 0.00027 | 33950 |
| 2457276.36680 | 0.00018 | 36424 |
| 2457291.38805 | 0.00021 | 36525 |
| 2457594.48734 | 0.00024 | 38563 |
| 2457609.50881 | 0.00016 | 38664 |
| 2458038.42837 | 0.00022 | 41548 |
| 2458039.32057 | 0.00017 | 41554 |
| 2458042.29484 | 0.00019 | 41574 |
| 2458385.40088 | 0.00027 | 43881 |
| 2458386.44190 | 0.00004 | 43888 |
| 2458719.43614 | 0.00012 | 46127 |
| 2458721.36860 | 0.00023 | 46140 |
| 2458784.42820 | 0.00020 | 46564 |
| 2458806.43966 | 0.00018 | 46712 |
| 2458911.43828 | 0.00015 | 47418 |
| 2458925.41820 | 0.00014 | 47512 |
| 2459041.42251 | 0.00017 | 48292 |
| 2459053.46929 | 0.00025 | 48373 |
| 2459056.44407 | 0.00010 | 48393 |
| 2459110.43117 | 0.00022 | 48756 |
| 2459117.42066 | 0.00011 | 48803 |
| 2459389.43726 | 0.00014 | 50632 |
| 2459414.42257 | 0.00010 | 50800 |
| 2459415.46405 | 0.00014 | 50807 |

Table 3.8. Eclipse times, errors and cycle numbers for HS 0220+0603 observed and measured by the author in this study.

| *Eclipse time (HJD)* | *Error (d)* | *Cycle Number* |
|---|---|---|
| 2454061.32109 | 0.00048 | 10038 |
| 2454081.31479 | 0.00032 | 10172 |
| 2454081.46403 | 0.00018 | 10173 |
| 2454086.38783 | 0.00026 | 10206 |
| 2455156.35608 | 0.00028 | 17377 |
| 2455188.43603 | 0.00027 | 17592 |
| 2455200.37262 | 0.00034 | 17672 |
| 2455490.43180 | 0.00028 | 19616 |
| 2455515.34977 | 0.00031 | 19783 |
| 2455533.40410 | 0.00029 | 19904 |
| 2455867.48013 | 0.00024 | 22143 |
| 2455884.48964 | 0.00012 | 22257 |
| 2456249.45127 | 0.00022 | 24703 |
| 2456250.49598 | 0.00015 | 24710 |
| 2456266.46118 | 0.00022 | 24817 |
| 2456609.34044 | 0.00042 | 27115 |
| 2456619.33720 | 0.00028 | 27182 |
| 2456955.50247 | 0.00033 | 29435 |
| 2456985.34355 | 0.00024 | 29635 |
| 2457354.48328 | 0.00027 | 32109 |
| 2457403.27389 | 0.00013 | 32436 |
| 2457407.30240 | 0.00016 | 32463 |
| 2457684.38159 | 0.00023 | 34320 |
| 2457698.40661 | 0.00021 | 34414 |
| 2458054.41601 | 0.00022 | 36800 |
| 2458082.31770 | 0.00022 | 36987 |
| 2458477.27022 | 0.00022 | 39634 |
| 2458492.34036 | 0.00031 | 39735 |
| 2458817.46349 | 0.00020 | 41914 |
| 2458819.40361 | 0.00011 | 41927 |
| 2458822.38782 | 0.00018 | 41947 |
| 2459158.40349 | 0.00019 | 44199 |
| 2459176.45724 | 0.00022 | 44320 |
| 2459189.43843 | 0.00018 | 44407 |
| 2459584.39131 | 0.00034 | 47054 |
| 2459597.37209 | 0.00038 | 47141 |
| 2459870.42209 | 0.00032 | 48971 |

Table 3.9. Eclipse times, errors and cycle numbers for BP Lyn observed and measured by the author in this study.

| *Eclipse time (HJD)* | *Error (d)* | *Cycle Number* |
|---|---|---|
| 2454186.44462 | 0.00069 | 41257 |
| 2454891.36892 | 0.00095 | 45870 |
| 2454906.49781 | 0.00084 | 45969 |
| 2455239.32473 | 0.00058 | 48147 |
| 2455260.41122 | 0.00042 | 48285 |
| 2455263.31415 | 0.00049 | 48304 |
| 2455571.38461 | 0.00074 | 50320 |
| 2455594.30701 | 0.00042 | 50470 |
| 2455619.52087 | 0.00059 | 50635 |
| 2455914.44759 | 0.00041 | 52565 |
| 2455930.34125 | 0.00063 | 52669 |
| 2455932.32762 | 0.00066 | 52682 |
| 2455942.41314 | 0.00039 | 52748 |
| 2455991.31277 | 0.00055 | 53068 |
| 2456016.37415 | 0.00052 | 53232 |
| 2456338.34928 | 0.00069 | 55339 |
| 2456343.39349 | 0.00056 | 55372 |
| 2456355.31121 | 0.00039 | 55450 |
| 2456356.38067 | 0.00034 | 55457 |
| 2456410.47808 | 0.00063 | 55811 |
| 2456415.36759 | 0.00065 | 55843 |
| 2456684.31780 | 0.00044 | 57603 |
| 2456728.32764 | 0.00126 | 57891 |
| 2457021.42236 | 0.00055 | 59809 |
| 2457059.32139 | 0.00083 | 60057 |
| 2457062.37785 | 0.00044 | 60077 |
| 2457433.40551 | 0.00051 | 62505 |
| 2457447.31156 | 0.00062 | 62596 |
| 2457455.41132 | 0.00026 | 62649 |
| 2457758.43751 | 0.00046 | 64632 |
| 2457778.30406 | 0.00037 | 64762 |
| 2458137.41551 | 0.00031 | 67112 |
| 2458161.40539 | 0.00031 | 67269 |
| 2458162.32198 | 0.00042 | 67275 |
| 2458163.39102 | 0.00040 | 67282 |
| 2458514.40265 | 0.00055 | 69579 |
| 2458517.30581 | 0.00051 | 69598 |
| 2458526.32134 | 0.00072 | 69657 |
| 2458539.31249 | 0.00049 | 69742 |
| 2458864.34320 | 0.00073 | 71869 |
| 2458886.34755 | 0.00040 | 72013 |
| 2458925.46891 | 0.00045 | 72269 |
| 2459240.41430 | 0.00037 | 74330 |
| 2459258.44656 | 0.00062 | 74448 |
| 2459271.43627 | 0.00051 | 74533 |

Table 3.10. Eclipse times, errors and cycle numbers for BH Lyn observed and measured by the author in this study.

| *Eclipse time (HJD)* | *Error (d)* | *Cycle Number* |
|---|---|---|
| 2454181.48914 | 0.00029 | 44915 |
| 2454186.32132 | 0.00042 | 44946 |
| 2454199.41436 | 0.00053 | 45030 |
| 2454482.32954 | 0.00048 | 46845 |
| 2454834.45234 | 0.00046 | 49104 |
| 2454884.33284 | 0.00052 | 49424 |
| 2455247.36666 | 0.00027 | 51753 |
| 2455260.46000 | 0.00033 | 51837 |
| 2455267.31793 | 0.00059 | 51881 |
| 2455594.34608 | 0.00035 | 53979 |
| 2455628.32676 | 0.00041 | 54197 |
| 2455670.41251 | 0.00040 | 54467 |
| 2455675.40111 | 0.00031 | 54499 |
| 2455895.34197 | 0.00038 | 55910 |
| 2455902.35570 | 0.00039 | 55955 |
| 2455941.32605 | 0.00040 | 56205 |
| 2455992.45237 | 0.00076 | 56533 |
| 2455994.32276 | 0.00053 | 56545 |
| 2455994.47949 | 0.00087 | 56546 |
| 2456028.45927 | 0.00021 | 56764 |
| 2456298.43632 | 0.00040 | 58496 |
| 2456356.42123 | 0.00032 | 58868 |
| 2456382.45272 | 0.00022 | 59035 |
| 2456699.34816 | 0.00027 | 61068 |
| 2456707.45398 | 0.00027 | 61120 |
| 2456726.47051 | 0.00038 | 61242 |
| 2457017.33447 | 0.00048 | 63108 |
| 2457020.45224 | 0.00043 | 63128 |
| 2457021.38791 | 0.00056 | 63134 |
| 2457433.36610 | 0.00032 | 65777 |
| 2457443.34252 | 0.00031 | 65841 |
| 2457460.48838 | 0.00021 | 65951 |
| 2457721.42424 | 0.00040 | 67625 |
| 2457727.34740 | 0.00043 | 67663 |
| 2458155.38234 | 0.00031 | 70409 |
| 2458163.33224 | 0.00019 | 70460 |
| 2458172.37306 | 0.00035 | 70518 |
| 2458840.45547 | 0.00035 | 74804 |
| 2458864.30434 | 0.00028 | 74957 |
| 2458868.35774 | 0.00038 | 74983 |
| 2459221.41548 | 0.00019 | 77248 |
| 2459238.40565 | 0.00017 | 77357 |
| 2459256.33224 | 0.00016 | 77472 |

Table 3.11. Eclipse times, errors and cycle numbers for LX Ser observed and measured by the author in this study.

| *Eclipse time (HJD)* | *Error (d)* | *Cycle Number* |
|---|---|---|
| 2454316.41420 | 0.00032 | 63266 |
| 2454628.52570 | 0.00023 | 65236 |
| 2454976.44297 | 0.00038 | 67432 |
| 2454994.50414 | 0.00026 | 67546 |
| 2455001.47525 | 0.00033 | 67590 |
| 2455037.43960 | 0.00020 | 67817 |
| 2455662.45627 | 0.00040 | 71762 |
| 2455663.40637 | 0.00045 | 71768 |
| 2455672.43730 | 0.00041 | 71825 |
| 2455778.42860 | 0.00031 | 72494 |
| 2456028.43528 | 0.00029 | 74072 |
| 2456076.44023 | 0.00042 | 74375 |
| 2456088.48102 | 0.00025 | 74451 |
| 2456384.43183 | 0.00028 | 76319 |
| 2456403.44388 | 0.00026 | 76439 |
| 2456410.41513 | 0.00047 | 76483 |
| 2456412.47433 | 0.00021 | 76496 |
| 2456782.41518 | 0.00010 | 78831 |
| 2456792.39591 | 0.00050 | 78894 |
| 2456798.41635 | 0.00018 | 78932 |
| 2457134.45163 | 0.00014 | 81053 |
| 2457159.48411 | 0.00019 | 81211 |
| 2457163.44478 | 0.00017 | 81236 |
| 2457491.40048 | 0.00055 | 83306 |
| 2457496.47045 | 0.00039 | 83338 |
| 2457506.45164 | 0.00033 | 83401 |
| 2457900.47370 | 0.00029 | 85888 |
| 2457901.42457 | 0.00029 | 85894 |
| 2457939.44889 | 0.00027 | 86134 |
| 2458227.47854 | 0.00022 | 87952 |
| 2458228.42935 | 0.00028 | 87958 |
| 2458241.42094 | 0.00033 | 88040 |
| 2458246.49055 | 0.00023 | 88072 |
| 2458593.45853 | 0.00026 | 90262 |
| 2458594.40834 | 0.00031 | 90268 |
| 2458599.47913 | 0.00019 | 90300 |
| 2458603.43930 | 0.00015 | 90325 |
| 2458943.43549 | 0.00035 | 92471 |
| 2458946.44578 | 0.00033 | 92490 |
| 2458949.45551 | 0.00039 | 92509 |
| 2459341.41767 | 0.00021 | 94983 |
| 2459350.44820 | 0.00011 | 95040 |
| 2459354.40902 | 0.00033 | 95065 |
| 2459704.38657 | 0.00024 | 97274 |
| 2459713.41715 | 0.00044 | 97331 |
| 2459744.47026 | 0.00029 | 97527 |

Table 3.12. Eclipse times, errors and cycle numbers for UU Aqr observed and measured by the author in this study.

| *Eclipse time (HJD)* | *Error (d)* | *Cycle Number* |
|---|---|---|
| 2454323.44995 | 0.00046 | 48760 |
| 2454357.47405 | 0.00027 | 48968 |
| 2454365.48955 | 0.00036 | 49017 |
| 2454728.47437 | 0.00051 | 51236 |
| 2454735.34486 | 0.00034 | 51278 |
| 2454736.32601 | 0.00056 | 51284 |
| 2454789.32574 | 0.00032 | 51608 |
| 2455038.45994 | 0.00069 | 53131 |
| 2455059.39716 | 0.00052 | 53259 |
| 2455106.34585 | 0.00043 | 53546 |
| 2455469.49424 | 0.00052 | 55766 |
| 2455490.26865 | 0.00048 | 55893 |
| 2455778.49715 | 0.00019 | 57655 |
| 2455795.50952 | 0.00019 | 57759 |
| 2455893.33048 | 0.00019 | 58357 |
| 2456159.47572 | 0.00030 | 59984 |
| 2456160.45716 | 0.00033 | 59990 |
| 2456162.42044 | 0.00044 | 60002 |
| 2456215.42071 | 0.00018 | 60326 |
| 2456512.48351 | 0.00045 | 62142 |
| 2456523.44298 | 0.00024 | 62209 |
| 2456532.43936 | 0.00066 | 62264 |
| 2456611.28481 | 0.00045 | 62746 |
| 2456612.26681 | 0.00033 | 62752 |
| 2456893.46177 | 0.00016 | 64471 |
| 2456903.44070 | 0.00038 | 64532 |
| 2456904.42104 | 0.00036 | 64538 |
| 2457258.40900 | 0.00022 | 66702 |
| 2457262.49817 | 0.00020 | 66727 |
| 2457275.42161 | 0.00012 | 66806 |
| 2457609.45204 | 0.00021 | 68848 |
| 2457617.46778 | 0.00040 | 68897 |
| 2457642.49568 | 0.00040 | 69050 |
| 2457979.47132 | 0.00035 | 71110 |
| 2457989.44973 | 0.00043 | 71171 |
| 2457993.37551 | 0.00026 | 71195 |
| 2458352.43417 | 0.00030 | 73390 |
| 2458360.44924 | 0.00025 | 73439 |
| 2458362.41308 | 0.00018 | 73451 |
| 2458363.39380 | 0.00024 | 73457 |
| 2458766.45546 | 0.00027 | 75921 |
| 2458784.28619 | 0.00018 | 76030 |
| 2458799.33537 | 0.00016 | 76122 |
| 2459102.44929 | 0.00024 | 77975 |
| 2459106.37518 | 0.00017 | 77999 |
| 2459107.35642 | 0.00036 | 78005 |
| 2459476.39397 | 0.00025 | 80261 |
| 2459478.35668 | 0.00047 | 80273 |
| 2459498.31321 | 0.00043 | 80395 |
| 2459499.29486 | 0.00029 | 80401 |
| 2459799.46523 | 0.00030 | 82236 |
| 2459859.33507 | 0.00055 | 82602 |
| 2459902.35724 | 0.00014 | 82865 |

Table 3.13. Eclipse times, errors and cycle numbers for V1776 Cyg observed and measured by the author in this study.

| Eclipse time (HJD) | Error (d) | Cycle Number |
|---|---|---|
| 2454238.48406 | 0.00059 | 43643 |
| 2454254.46252 | 0.00044 | 43740 |
| 2454306.51977 | 0.00050 | 44056 |
| 2454314.42730 | 0.00053 | 44104 |
| 2454646.54029 | 0.00092 | 46120 |
| 2454668.44971 | 0.00092 | 46253 |
| 2454670.42804 | 0.00080 | 46265 |
| 2454770.42363 | 0.00115 | 46872 |
| 2454994.46940 | 0.00068 | 48232 |
| 2455037.46488 | 0.00052 | 48493 |
| 2455057.39969 | 0.00051 | 48614 |
| 2455176.34096 | 0.00062 | 49336 |
| 2455460.34923 | 0.00100 | 51060 |
| 2455494.45030 | 0.00101 | 51267 |
| 2455778.46040 | 0.00052 | 52991 |
| 2455849.46194 | 0.00052 | 53422 |
| 2455893.28160 | 0.00088 | 53688 |
| 2456132.48198 | 0.00094 | 55140 |
| 2456144.51030 | 0.00076 | 55213 |
| 2456150.43908 | 0.00044 | 55249 |
| 2456160.48760 | 0.00054 | 55310 |
| 2456176.46818 | 0.00069 | 55407 |
| 2456445.48448 | 0.00041 | 57040 |
| 2456446.47452 | 0.00053 | 57046 |
| 2456450.42722 | 0.00071 | 57070 |
| 2456506.43829 | 0.00052 | 57410 |
| 2456803.46380 | 0.00068 | 59213 |
| 2456834.43307 | 0.00149 | 59401 |
| 2456840.52859 | 0.00068 | 59438 |
| 2456842.50564 | 0.00072 | 59450 |
| 2456893.40936 | 0.00044 | 59759 |
| 2457172.47754 | 0.00064 | 61453 |
| 2457174.45542 | 0.00088 | 61465 |
| 2457177.42075 | 0.00094 | 61483 |
| 2457532.43120 | 0.00068 | 63638 |
| 2457533.41877 | 0.00098 | 63644 |
| 2457545.44595 | 0.00083 | 63717 |
| 2457959.43351 | 0.00019 | 66230 |
| 2457971.46058 | 0.00068 | 66303 |
| 2457976.40197 | 0.00062 | 66333 |
| 2458246.40926 | 0.00048 | 67972 |
| 2458255.46985 | 0.00082 | 68027 |
| 2458272.43802 | 0.00051 | 68130 |
| 2458284.46341 | 0.00060 | 68203 |
| 2458643.42871 | 0.00069 | 70382 |
| 2458655.45529 | 0.00056 | 70455 |
| 2458656.44378 | 0.00057 | 70461 |
| 2458667.48111 | 0.00045 | 70528 |
| 2458983.44913 | 0.00029 | 72446 |
| 2458995.47540 | 0.00048 | 72519 |
| 2458997.45364 | 0.00043 | 72531 |
| 2458998.44053 | 0.00048 | 72537 |
| 2459106.34437 | 0.00060 | 73192 |
| 2459112.43907 | 0.00056 | 73229 |
| 2459113.42785 | 0.00089 | 73235 |
| 2459366.46757 | 0.00082 | 74771 |
| 2459367.45497 | 0.00075 | 74777 |
| 2459369.43276 | 0.00032 | 74789 |

Table 3.14. Eclipse times, errors and cycle numbers for RW Tri observed and measured by the author in this study.

| Eclipse time (HJD) | Error (d) | Cycle Number |
|---|---|---|
| 2454392.38737 | 0.00024 | 57197 |
| 2454419.51756 | 0.00027 | 57314 |
| 2454447.34346 | 0.00020 | 57434 |
| 2454789.37226 | 0.00041 | 58909 |
| 2454810.47333 | 0.00064 | 59000 |
| 2454835.28542 | 0.00050 | 59107 |
| 2455063.45767 | 0.00047 | 60091 |
| 2455106.35664 | 0.00047 | 60276 |
| 2455172.44338 | 0.00026 | 60561 |
| 2455487.34152 | 0.00042 | 61919 |
| 2455490.35562 | 0.00017 | 61932 |
| 2455533.48590 | 0.00023 | 62118 |
| 2455822.41233 | 0.00026 | 63364 |
| 2455828.44141 | 0.00023 | 63390 |
| 2455867.39741 | 0.00048 | 63558 |
| 2455881.31079 | 0.00014 | 63618 |
| 2455889.42621 | 0.00028 | 63653 |
| 2455914.23796 | 0.00028 | 63760 |
| 2455950.41154 | 0.00024 | 63916 |
| 2455953.42610 | 0.00051 | 63929 |
| 2455957.36910 | 0.00018 | 63946 |
| 2456200.38189 | 0.00029 | 64994 |
| 2456215.45437 | 0.00053 | 65059 |
| 2456228.43987 | 0.00027 | 65115 |
| 2456609.42450 | 0.00022 | 66758 |
| 2456619.39553 | 0.00012 | 66801 |
| 2456636.32322 | 0.00019 | 66874 |
| 2456922.46690 | 0.00018 | 68108 |
| 2456933.36541 | 0.00036 | 68155 |
| 2456935.45247 | 0.00017 | 68164 |
| 2457320.37908 | 0.00023 | 69824 |
| 2457327.33548 | 0.00025 | 69854 |
| 2457403.39345 | 0.00023 | 70182 |
| 2457623.45120 | 0.00056 | 71131 |
| 2457642.46534 | 0.00016 | 71213 |
| 2457645.47976 | 0.00013 | 71226 |
| 2458054.29026 | 0.00041 | 72989 |
| 2458059.39146 | 0.00026 | 73011 |
| 2458062.40634 | 0.00039 | 73024 |
| 2458379.39042 | 0.00036 | 74391 |
| 2458398.40463 | 0.00018 | 74473 |
| 2458401.41934 | 0.00037 | 74486 |
| 2458784.48887 | 0.00025 | 76138 |
| 2458817.41674 | 0.00013 | 76280 |
| 2458822.28650 | 0.00020 | 76301 |
| 2458827.38796 | 0.00029 | 76323 |
| 2459101.47488 | 0.00031 | 77505 |
| 2459114.45966 | 0.00030 | 77561 |
| 2459157.35817 | 0.00028 | 77746 |
| 2459221.35700 | 0.00029 | 78022 |
| 2459236.42944 | 0.00053 | 78087 |
| 2459273.29920 | 0.00025 | 78246 |
| 2459521.41392 | 0.00024 | 79316 |
| 2459541.35608 | 0.00023 | 79402 |
| 2459580.31229 | 0.00045 | 79570 |
| 2459912.36734 | 0.00021 | 81002 |
| 2459921.41077 | 0.00022 | 81041 |
| 2459928.36699 | 0.00022 | 81071 |

Table 3.15. Eclipse times, errors and cycle numbers for 1RXS J064434.5+334451 observed and measured by the author in this study.

| Eclipse time (HJD) | Error (d) | Cycle Number |
|---|---|---|
| 2455307.42924 | 0.00074 | 7067 |
| 2455310.39210 | 0.00056 | 7078 |
| 2455313.35557 | 0.00049 | 7089 |
| 2455627.44814 | 0.00048 | 8255 |
| 2455629.33392 | 0.00043 | 8262 |
| 2455634.45149 | 0.00035 | 8281 |
| 2455655.46296 | 0.00045 | 8359 |
| 2455658.42635 | 0.00025 | 8370 |
| 2455682.39947 | 0.00042 | 8459 |
| 2455685.36351 | 0.00051 | 8470 |
| 2455850.48993 | 0.00045 | 9083 |
| 2455854.53082 | 0.00023 | 9098 |
| 2455891.43482 | 0.00015 | 9235 |
| 2455905.44214 | 0.00063 | 9287 |
| 2455914.33106 | 0.00043 | 9320 |
| 2455924.29847 | 0.00046 | 9357 |
| 2455932.37955 | 0.00032 | 9387 |
| 2455949.35041 | 0.00027 | 9450 |
| 2455953.38926 | 0.00037 | 9465 |
| 2455957.43085 | 0.00052 | 9480 |
| 2455959.31737 | 0.00024 | 9487 |
| 2455960.39430 | 0.00028 | 9491 |
| 2455991.37304 | 0.00060 | 9606 |
| 2455998.37693 | 0.00019 | 9632 |
| 2456006.45817 | 0.00040 | 9662 |
| 2456012.38313 | 0.00029 | 9684 |
| 2456013.46042 | 0.00017 | 9688 |
| 2456029.35457 | 0.00024 | 9747 |
| 2456267.47943 | 0.00039 | 10631 |
| 2456274.48380 | 0.00059 | 10657 |
| 2456294.41725 | 0.00019 | 10731 |
| 2456338.32439 | 0.00028 | 10894 |
| 2456341.28762 | 0.00025 | 10905 |
| 2456343.44229 | 0.00024 | 10913 |
| 2456382.50149 | 0.00026 | 11058 |
| 2456384.38677 | 0.00037 | 11065 |
| 2456398.39477 | 0.00022 | 11117 |
| 2456655.37919 | 0.00015 | 12071 |
| 2456662.38307 | 0.00019 | 12097 |
| 2456677.46827 | 0.00025 | 12153 |
| 2456994.52257 | 0.00036 | 13330 |
| 2457000.44891 | 0.00030 | 13352 |
| 2457016.34157 | 0.00023 | 13411 |
| 2457042.47208 | 0.00021 | 13508 |
| 2457045.43393 | 0.00030 | 13519 |
| 2457047.31894 | 0.00015 | 13526 |
| 2457104.42741 | 0.00020 | 13738 |
| 2457402.35525 | 0.00028 | 14844 |
| 2457407.47333 | 0.00029 | 14863 |
| 2457408.28185 | 0.00047 | 14866 |
| 2457702.43865 | 0.00050 | 15958 |
| 2457723.45002 | 0.00031 | 16036 |
| 2457726.41335 | 0.00022 | 16047 |
| 2458074.44349 | 0.00019 | 17339 |
| 2458085.48724 | 0.00047 | 17380 |
| 2458161.45098 | 0.00026 | 17662 |
| 2458477.42648 | 0.00035 | 18835 |
| 2458493.31897 | 0.00048 | 18894 |
| 2458498.43788 | 0.00023 | 18913 |
| 2458827.34416 | 0.00065 | 20134 |
| 2458855.35853 | 0.00026 | 20238 |
| 2458866.40335 | 0.00025 | 20279 |
| 2459189.38307 | 0.00018 | 21478 |

*Table 3.15 continued on next page.*

Table 3.15. Eclipse times, errors and cycle numbers for 1RXS J064434.5+334451 observed and measured by the author in this study, cont.

| *Eclipse time (HJD)* | *Error (d)* | *Cycle Number* |
|---|---|---|
| 2459196.38696 | 0.00029 | 21504 |
| 2459203.39060 | 0.00036 | 21530 |
| 2459592.36698 | 0.00064 | 22974 |
| 2459593.44501 | 0.00024 | 22978 |
| 2459596.40755 | 0.00022 | 22989 |
| 2459995.35223 | 0.00017 | 24470 |
| 2460002.35630 | 0.00027 | 24496 |

Table 3.16. Eclipse times, errors and cycle numbers for AC Cnc observed and measured by the author in this study.

| *Eclipse time (HJD)* | *Error (d)* | *Cycle Number* |
|---|---|---|
| 2454199.45197 | 0.00026 | 32978 |
| 2454507.44198 | 0.00021 | 34003 |
| 2454891.45161 | 0.00036 | 35281 |
| 2454892.35306 | 0.00032 | 35284 |
| 2455260.43835 | 0.00023 | 36509 |
| 2455270.35440 | 0.00042 | 36542 |
| 2455619.50814 | 0.00082 | 37704 |
| 2455630.32565 | 0.00024 | 37740 |
| 2455675.39674 | 0.00047 | 37890 |
| 2455949.43118 | 0.00029 | 38802 |
| 2455959.34723 | 0.00034 | 38835 |
| 2455983.38539 | 0.00020 | 38915 |
| 2455994.50332 | 0.00107 | 38952 |
| 2455998.41002 | 0.00037 | 38965 |
| 2456001.41362 | 0.00048 | 38975 |
| 2456308.50261 | 0.00049 | 39997 |
| 2456330.43727 | 0.00025 | 40070 |
| 2456342.45628 | 0.00030 | 40110 |
| 2456680.49280 | 0.00037 | 41235 |
| 2456684.39897 | 0.00051 | 41248 |
| 2456699.42304 | 0.00065 | 41298 |
| 2457047.37473 | 0.00043 | 42456 |
| 2457059.39437 | 0.00051 | 42496 |
| 2457080.42748 | 0.00020 | 42566 |
| 2457421.46916 | 0.00028 | 43701 |
| 2457430.48357 | 0.00023 | 43731 |
| 2457433.48793 | 0.00025 | 43741 |
| 2457763.41179 | 0.00053 | 44839 |
| 2457803.37541 | 0.00022 | 44972 |
| 2457815.39490 | 0.00032 | 45012 |
| 2457827.41406 | 0.00027 | 45052 |
| 2458125.48746 | 0.00048 | 46044 |
| 2458137.50672 | 0.00026 | 46084 |
| 2458162.44626 | 0.00024 | 46167 |
| 2458519.41331 | 0.00030 | 47355 |
| 2458537.44254 | 0.00024 | 47415 |
| 2458568.39100 | 0.00031 | 47518 |
| 2458595.43418 | 0.00047 | 47608 |
| 2458869.46939 | 0.00025 | 48520 |
| 2458910.33507 | 0.00062 | 48656 |
| 2458925.35767 | 0.00019 | 48706 |
| 2459256.48343 | 0.00025 | 49808 |
| 2459272.40883 | 0.00031 | 49861 |
| 2459281.42325 | 0.00020 | 49891 |
| 2459632.37981 | 0.00047 | 51059 |
| 2459659.42252 | 0.00019 | 51149 |
| 2459665.43239 | 0.00029 | 51169 |
| 2459989.34592 | 0.00057 | 52247 |
| 2460001.36523 | 0.00027 | 52287 |

Table 3.17. Eclipse times, errors and cycle numbers for V363 Aur observed and measured by the author in this study.

| *Eclipse time (HJD)* | *Error (d)* | *Cycle Number* |
|---|---|---|
| 2454181.39163 | 0.00043 | 29957 |
| 2454392.44674 | 0.00017 | 30614 |
| 2454447.37885 | 0.00024 | 30785 |
| 2454471.47221 | 0.00031 | 30860 |
| 2454473.39980 | 0.00037 | 30866 |
| 2454810.38137 | 0.00031 | 31915 |
| 2454827.40653 | 0.00042 | 31968 |
| 2454835.43772 | 0.00044 | 31993 |
| 2454891.33360 | 0.00054 | 32167 |
| 2454892.29747 | 0.00021 | 32170 |
| 2455188.48144 | 0.00054 | 33092 |
| 2455191.37255 | 0.00040 | 33101 |
| 2455200.36736 | 0.00026 | 33129 |
| 2455515.50429 | 0.00013 | 34110 |
| 2455516.46885 | 0.00021 | 34113 |
| 2455524.49896 | 0.00034 | 34138 |
| 2455526.42586 | 0.00026 | 34144 |
| 2455627.29626 | 0.00020 | 34458 |
| 2455634.36298 | 0.00020 | 34480 |
| 2455649.46157 | 0.00047 | 34527 |
| 2455854.41351 | 0.00026 | 35165 |
| 2455888.46463 | 0.00016 | 35271 |
| 2455891.35618 | 0.00021 | 35280 |
| 2455905.49122 | 0.00039 | 35324 |
| 2455914.48560 | 0.00015 | 35352 |
| 2455950.46438 | 0.00013 | 35464 |
| 2455954.31900 | 0.00028 | 35476 |
| 2455994.47452 | 0.00029 | 35601 |
| 2456014.39134 | 0.00019 | 35663 |
| 2456215.48745 | 0.00015 | 36289 |
| 2456262.38905 | 0.00051 | 36435 |
| 2456291.30073 | 0.00025 | 36525 |
| 2456344.30534 | 0.00025 | 36690 |
| 2456655.26652 | 0.00034 | 37658 |
| 2456662.33389 | 0.00019 | 37680 |
| 2456677.43090 | 0.00025 | 37727 |
| 2456698.31132 | 0.00043 | 37792 |
| 2456707.30608 | 0.00013 | 37820 |
| 2456952.41251 | 0.00024 | 38583 |
| 2456985.50008 | 0.00022 | 38686 |
| 2456994.49490 | 0.00042 | 38714 |
| 2457349.46509 | 0.00025 | 39819 |
| 2457377.41312 | 0.00026 | 39906 |
| 2457429.45427 | 0.00015 | 40068 |
| 2457721.46196 | 0.00022 | 40977 |
| 2457741.37878 | 0.00022 | 41039 |
| 2457846.42357 | 0.00028 | 41366 |
| 2458066.47383 | 0.00023 | 42051 |
| 2458074.50509 | 0.00036 | 42076 |
| 2458085.42672 | 0.00020 | 42110 |
| 2458441.36127 | 0.00025 | 43218 |
| 2458465.45436 | 0.00018 | 43293 |
| 2458819.46136 | 0.00016 | 44395 |
| 2458822.35212 | 0.00017 | 44404 |
| 2458868.28933 | 0.00028 | 44547 |
| 2459148.41099 | 0.00018 | 45419 |
| 2459157.40619 | 0.00036 | 45447 |
| 2459164.47295 | 0.00026 | 45469 |
| 2459273.37328 | 0.00021 | 45808 |
| 2459575.34002 | 0.00015 | 46748 |
| 2459584.33471 | 0.00016 | 46776 |
| 2459975.28343 | 0.00019 | 47993 |

Table 3.18. Eclipse times, errors and cycle numbers for BT Mon observed and measured by the author in this study.

| *Eclipse time (HJD)* | *Error (d)* | *Cycle Number* |
|---|---|---|
| 2454447.47617 | 0.00050 | 32820 |
| 2454891.44778 | 0.00061 | 34150 |
| 2454892.44988 | 0.00053 | 34153 |
| 2455238.27878 | 0.00058 | 35189 |
| 2455239.28089 | 0.00095 | 35192 |
| 2455257.30609 | 0.00040 | 35246 |
| 2455260.31093 | 0.00048 | 35255 |
| 2455277.33531 | 0.00079 | 35306 |
| 2455571.42510 | 0.00067 | 36187 |
| 2455595.46030 | 0.00073 | 36259 |
| 2455600.46698 | 0.00109 | 36274 |
| 2455619.49354 | 0.00056 | 36331 |
| 2455960.31808 | 0.00104 | 37352 |
| 2455968.33013 | 0.00055 | 37376 |
| 2455987.35688 | 0.00064 | 37433 |
| 2455992.36366 | 0.00082 | 37448 |
| 2456001.37745 | 0.00090 | 37475 |
| 2456011.39153 | 0.00070 | 37505 |
| 2456294.46579 | 0.00045 | 38353 |
| 2456330.51701 | 0.00047 | 38461 |
| 2456338.52784 | 0.00048 | 38485 |
| 2456684.35905 | 0.00079 | 39521 |
| 2456707.39142 | 0.00025 | 39590 |
| 2456725.41702 | 0.00037 | 39644 |
| 2457011.49426 | 0.00023 | 40501 |
| 2457017.50343 | 0.00037 | 40519 |
| 2457020.50843 | 0.00037 | 40528 |
| 2457395.37986 | 0.00051 | 41651 |
| 2457402.39016 | 0.00052 | 41672 |
| 2457407.39767 | 0.00062 | 41687 |
| 2457803.29994 | 0.00042 | 42873 |
| 2457815.31701 | 0.00067 | 42909 |
| 2457827.33521 | 0.00038 | 42945 |
| 2457828.33500 | 0.00044 | 42948 |
| 2458137.44694 | 0.00033 | 43874 |
| 2458151.46592 | 0.00045 | 43916 |
| 2458161.48037 | 0.00039 | 43946 |
| 2458529.34362 | 0.00053 | 45048 |
| 2458536.35358 | 0.00047 | 45069 |
| 2458537.35429 | 0.00056 | 45072 |
| 2458866.49427 | 0.00023 | 46058 |
| 2458869.49729 | 0.00039 | 46067 |
| 2459238.36139 | 0.00054 | 47172 |
| 2459249.37680 | 0.00034 | 47205 |

Table 4. Eclipse times, errors and cycle numbers for LX Ser measured by the author from observations by Cook and Dvorak in the AAVSO International Database.

| *Eclipse time (HJD)* | *Error (d)* | *Cycle Number* | *Observer* | *Eclipse time (HJD)* | *Error (d)* | *Cycle Number* | *Observer* |
|---|---|---|---|---|---|---|---|
| 2452777.87523 | 0.00050 | 53555 | Cook | 2458192.94006 | 0.00020 | 87734 | Dvorak |
| 2452778.82598 | 0.00056 | 53561 | Cook | 2458193.89137 | 0.00042 | 87740 | Dvorak |
| 2452779.77652 | 0.00072 | 53567 | Cook | 2458220.82416 | 0.00025 | 87910 | Dvorak |
| 2452779.93474 | 0.00057 | 53568 | Cook | 2458227.79532 | 0.00032 | 87954 | Dvorak |
| 2452780.88542 | 0.00044 | 53574 | Cook | 2458233.81608 | 0.00042 | 87992 | Dvorak |
| 2452781.83604 | 0.00049 | 53580 | Cook | 2458239.83639 | 0.00045 | 88030 | Dvorak |
| 2452782.78676 | 0.00050 | 53586 | Cook | 2458242.84637 | 0.00036 | 88049 | Dvorak |
| 2452782.94528 | 0.00051 | 53587 | Cook | 2458272.63221 | 0.00031 | 88237 | Dvorak |
| 2452786.74760 | 0.00036 | 53611 | Cook | 2458589.65567 | 0.00034 | 90238 | Dvorak |
| 2452786.90593 | 0.00022 | 53612 | Cook | 2458966.72465 | 0.00027 | 92618 | Dvorak |
| 2452787.85672 | 0.00041 | 53618 | Cook | 2459271.86577 | 0.00043 | 94544 | Dvorak |
| 2457882.73016 | 0.00052 | 85776 | Dvorak | 2459358.68647 | 0.00030 | 95092 | Dvorak |
| 2457889.70010 | 0.00037 | 85820 | Dvorak | 2459363.59809 | 0.00021 | 95123 | Dvorak |
| 2457899.68152 | 0.00027 | 85883 | Dvorak | 2459364.70675 | 0.00017 | 95130 | Dvorak |
| 2458167.90800 | 0.00032 | 87576 | Dvorak | 2459375.63919 | 0.00029 | 95199 | Dvorak |
| 2458181.85027 | 0.00035 | 87664 | Dvorak | 2459624.85361 | 0.00032 | 96772 | Dvorak |
| 2458187.87047 | 0.00021 | 87702 | Dvorak | 2459625.96292 | 0.00050 | 96779 | Dvorak |
| 2458191.83129 | 0.00031 | 87727 | Dvorak | 2459744.47026 | 0.00029 | 97527 | Dvorak |

Table 5. Sources of published eclipse times.

| *Star Name* | *Sources of published eclipse times* |
|---|---|
| HS 0728+6738 = V482 Cam | Rodriguez-Gil *et al.* (2004) |
| SW Sex = PG 1012-029 | Penning *et al.* (1984), Ashoka *et al.* (1994), Dhillon *et al.* (1997), Groot *et al.* (2001), Fang *et al.* (2020), one issue of BVSOLJ |
| DW UMa = PG1030+590 | Shafter *et al.* (1988), Dhillon *et al.* (1994), Bíró (2000), Stanishev *et al.* (2004), Dhillon *et al.* (2013), Boyd *et al.* (2017) (including observations from contributors to the Centre for Backyard Astrophysics), several issues of IBVS, BVSOLJ, OEJV |
| HS 0129+2933 = TT Tri | Warren *et al.* (2006), Rodriguez-Gil *et al.* (2007), Han *et al.* (2018) |
| V1315 Aql | Downes *et al.* (1986), Dhillon *et al.* (1991), Rutten *et al.* (1992), Hellier (1996), Papadaki *et al.* (2009), Fang and Qian (2021), a series of eclipse times by Cook published in *Observed Minima Times of Eclipsing Binaries, No 10* (Baldwin and Samolyk 2005) |
| PX And = PG0027+260 | Hellier and Robinson (1994), Stanishev *et al.* (2002), Han *et al.* (2018), several issues of IBVS |
| HS 0455+8315 | Rodriguez-Gil *et al.* (2007) |
| HS 0220+0603 | Rodriguez-Gil *et al.* (2007) |
| BP Lyn = PG0859+415 | Grauer *et al.* (1994), Still (1996), Han *et al.* (2018) |
| BH Lyn = PG0818+513 | Dhillon *et al.* (1992), Hoard and Szkody (1997), Stanishev *et al.* (2006), several issues of OEJV |
| LX Ser = Stepanyan's Star | Horne (1980), Africano and Klimke (1981), Young *et al.* (1981), Rutten *et al.* (1992), Li (2017), several issues of IBVS, BVSOLJ and OEJV |
| UU Aqr | Baptista *et al.* (1994), Han *et al.* (2018), several issues of BVSOLJ, IBVS and OEJV |
| V1776 Cyg = Lanning 90 | Garnavich *et al.* (1990) |
| RW Tri | Walker (1963), Africano *et al.* (1978), Robinson *et al.* (1991), Rutten *et al.* (1992), Smak (1995), Subebikova (2020), several issues of IBVS, OEJV and BVSOLJ |
| 1RXS J064434.5+334451 | Sing *et al.* (2007), Green (2008), Hernandez Santisteban (2017), Shafter and Bautista (2021) |
| AC Cnc | Yamasaki *et al.* (1983), Schlegel *et al.* (1984), Zhang *et al.* (1987), Thoroughgood *et al.* (2004), Qian *et al.* (2007), Bruch (2022), several issues of OEJV and IBVS |
| V363 Aur = Lanning 10 | Horne *et al.* (1982), Schlegel *et al.* (1986), Rutten *et al.* (1992), Thoroughgood *et al.* (2004), one issue of BVSOLJ |
| BT Mon | Robinson *et al.* (1982), Seitter (1984), Smith *et al.* (1998) |

*IBVS = Information Bulletin on Variable Stars: https://konkoly.hu/ibvs/; BVSOLJ = Bulletin of the Variable Star Observers League in Japan: http://vsolj.cetus-net.org/; OEJV = Open European Journal on Variable Stars: https://oejv.physics.muni.cz/*

Table 6. Weighted linear ephemerides for each star computed with all available data except in the case of SW Sex where there was a large change around 2017 and separate ephemerides are given for before and after this change. E is the cycle number.

| *Star Name* | *Weighted Linear Ephemeris* |
|---|---|
| HS 0728+6738 = V482 Cam | 2452001.32754(8) + 0.133619431(2) * E |
| SW Sex = PG 1012-029 (up to 2017) | 2444339.6502(2) + 0.134938480(2) * E |
| SW Sex = PG 1012-029 (after 2017) | 2444339.689(2) + 0.13493809(2) * E |
| DW UMa = PG1030+590 | 2446229.00633(8) + 0.136606541(1) * E |
| HS 0129+2933 = TT Tri | 2452540.5335(2) + 0.139637390(7) * E |
| V1315 Aql | 2445902.8387(2) + 0.139689996(2) * E |
| PX And = PG0027+260 | 2449238.8368(2) + 0.146352742(4) * E |
| HS 0455+8315 | 2451859.2458(2) + 0.148723946(5) * E |
| HS 0220+0603 | 2452563.57441(9) + 0.149207655(3) * E |
| BP Lyn = PG0859+415 | 2447881.8572(4) + 0.152812554(7) * E |
| BH Lyn = PG0818+513 | 2447180.3343(2) + 0.155875642(3) * E |
| LX Ser = Stepanyan's Star | 2444293.0227(2) + 0.158432503(2) * E |
| UU Aqr | 2446347.2670(2) + 0.163580423(3) * E |
| V1776 Cyg = Lanning 90 | 2447048.7932(3) + 0.164738652(5) * E |
| RW Tri | 2441129.3634(4) + 0.231883245(5) * E |
| 1RXS J064434.5+334451 | 2453403.7611(3) + 0.26937438(2) * E |
| AC Cnc | 2444290.3103(3) + 0.300477307(7) * E |
| V363 Aur = Lanning 10 | 2444557.981(2) + 0.32124074(6) * E |
| BT Mon | 2443491.7225(9) + 0.33381330(2) * E |

Table 7. Weighted quadratic ephemerides and mean rates of period change for stars showing evidence of either an increasing or decreasing orbital period. E is the cycle number.

| *Star Name* | *Weighted Quadratic Ephemeris* | *Mean Rate of Period Change (msec/year)* |
|---|---|---|
| HS 0728+6738 = V482 Cam | 2452001.3273(1) + 0.133619451(8) * E – 3(1)$10^{-13}$ * E2 | –0.16(6) |
| DW UMa = PG1030+590 | 2446229.0069(2) + 0.136606520(7) * E + 1.7(5)$10^{-13}$ * E2 | 0.08(3) |
| HS 0129+2933 = TT Tri | 2452540.5309(2) + 0.13963764(2) * E – 4.2(3)$10^{-12}$ * E2 | –1.9(2) |
| V1315 Aql | 2445902.8408(1) + 0.139689913(4) * E + 6.8(3)$10^{-13}$ * E2 | 0.31(2) |
| PX And = PG0027+260 | 2449238.8366(2) + 0.14635275(1) * E – 1(1)$10^{-13}$ * E2 | –0.06(6) |
| HS 0455+8315 | 2451859.2476(5) + 0.14872382(3) * E + 1.9(5)$10^{-12}$ * E2 | 0.8(2) |
| HS 0220+0603 | 2452563.57406(7) + 0.149207716(7) * E – 1.4(2) $10^{-12}$ * E2 | –0.58(6) |
| BP Lyn = PG0859+415 | 2447881.8584(4) + 0.15281244(2) * E + 1.5(3)$10^{-12}$ * E2 | 0.6(1) |
| BH Lyn = PG0818+513 | 2447180.3331(2) + 0.155875697(3) * E – 5.51(3)$10^{-13}$ * E2 | –0.22(1) |
| UU Aqr | 2446347.2656(1) + 0.163580565(8) * E – 1.70(9)$10^{-12}$ * E2 | –0.66(4) |
| V1776 Cyg = Lanning 90 | 2447048.7928(3) + 0.16473869(2) * E – 5(2)$10^{-13}$ * E2 | –0.19(8) |
| 1RXS J064434.5+334451 | 2453403.7596(3) + 0.26937469(5) * E – 1.2(2)$10^{-11}$ * E2 | –2.8(5) |
| V363 Aur = Lanning 10 | 2444557.9493(5) + 0.32124275(2) * E – 3.05(2)$10^{-11}$ * E2 | –5.98(4) |
| BT Mon | 2443491.7162(4) + 0.33381392(1) * E – 1.127(8)$10^{-11}$ * E2 | –2.13(2) |

Table 8. Parameters of sinusoidal fits relative to a linear ephemeris.

| *Star Name* | *Period of Sinusoidal Variation (year)* | *Half Amplitude of Sinusoidal Cariation (sec)* |
|---|---|---|
| SW Sex = PG 1012-029 (up to 2017) | 33 (2) | 46 (6) |
| LX Ser = Stepanyan's Star | 13.5 (2) | 55 (4) |
| RW Tri up (up to 2018) | 44.3 (7) | 191 (7) |
| AC Cnc | 37.8 (7) | 134 (14) |

Table 9. Parameters of sinusoidal fits relative to a quadratic ephemeris.

| *Star Name* | *Period of Sinusoidal Variation (year)* | *Half Amplitude of Sinusoidal Cariation (sec)* |
|---|---|---|
| DW UMa = PG1030+590 | 14.4 (3) | 38 (1) |
| HS 0129+2933 = TT Tri | 13.0 (4) | 47 (6) |
| 1RXS J064434.5+334451 | 6.2 (2) | 87 (10) |

## 5. O–C diagrams

In these O–C diagrams, data from the published literature or derived from observations in the AAVSO International Database are shown in black while eclipse times measured by the author are shown in red. Linear ephemerides are shown dotted in black, quadratic ephemerides dotted in magenta, and sinusoidal fits dotted in green. The passing years are marked above each diagram. O–C diagrams with similar apparent behavior are grouped together. To achieve a degree of consistency between these diagrams, we have used the same scale on the O–C axis except where the range of the data is significantly larger. It is worth stating explicitly that including these fits in the O–C diagrams is a subjective exercise which yields parameters that can be quantified but does not imply a physical interpretation. The O–C diagrams are described in five groups.

*HS 0728+6738 = V482 Cam, PX And = PG0027+260, HS 0220+0603, BH Lyn = PG0818+513, V1776 Cyg = Lanning 90* These stars show predominantly linear behavior with weak evidence of decreasing orbital period. Their O–C diagrams with linear and quadratic ephemerides are shown in Figure 2 and parameters of the quadratic ephemerides are given in Table 7.

*HS 0129+2933 = TT Tri, UU Aqr, 1RXS J064434.5+334451, V363 Aur = Lanning 10, BT Mon* These stars show stronger evidence of decreasing orbital periods. Their O–C diagrams with linear and quadratic ephemerides are shown in Figure 3 and parameters of the quadratic ephemerides are given in Table 7.

*DW UMa = PG1030+590, V1315 Aql, HS0455+8315, BP Lyn = PG0859+415* These stars show evidence of increasing orbital periods. Their O–C diagrams with linear and quadratic ephemerides are shown in Figure 4 and parameters of the quadratic ephemerides are given in Table 7.

*SW Sex = PG 1012-029, LX Ser = Stepanyan's Star, RW Tri, AC Cnc* These stars show evidence of sinusoidal variation in their orbital periods relative to a linear ephemeris. Figure 5 shows their O–C diagrams with sinusoidal fits relative to a linear ephemeris and Table 8 gives parameters of these sinusoidal fits.

*Stars also showing evidence of more complex behavior* In addition to their behavior described above, DW UMa, HS0129+2933 and 1RXS J064434.5+334451 also show evidence of sinusoidal variation in their orbital periods relative to a quadratic ephemeris. Figure 6 shows their O–C diagrams with sinusoidal fits relative to a quadratic ephemeris. Table 9 gives parameters of these sinusoidal fits. DW UMa now appears to be diverging from this sinusoidal pattern.

Figure 5 shows that both SW Sex and RW Tri recently experienced large decreases in their orbital periods.

Prior to 2017 (cycle ~100000) the mean orbital period of SW Sex over the previous 37 years had been 0.134938480(2) d with relatively weak sinusoidal modulation. During 2017 this reduced to 0.13493809(2) d, a decrease of 34 msec and a proportional change of $-2.9 \times 10^{-6}$.

Prior to 2018 (cycle ~74000) the mean orbital period of RW Tri over the previous 15 years had been 0.231883411(6) d with relatively strong sinusoidal modulation. Within a few months this changed and the mean orbital period since 2018 has been 0.23188288(6) d, a decrease of 46 msec and a proportional change of $-2.3 \times 10^{-6}$.

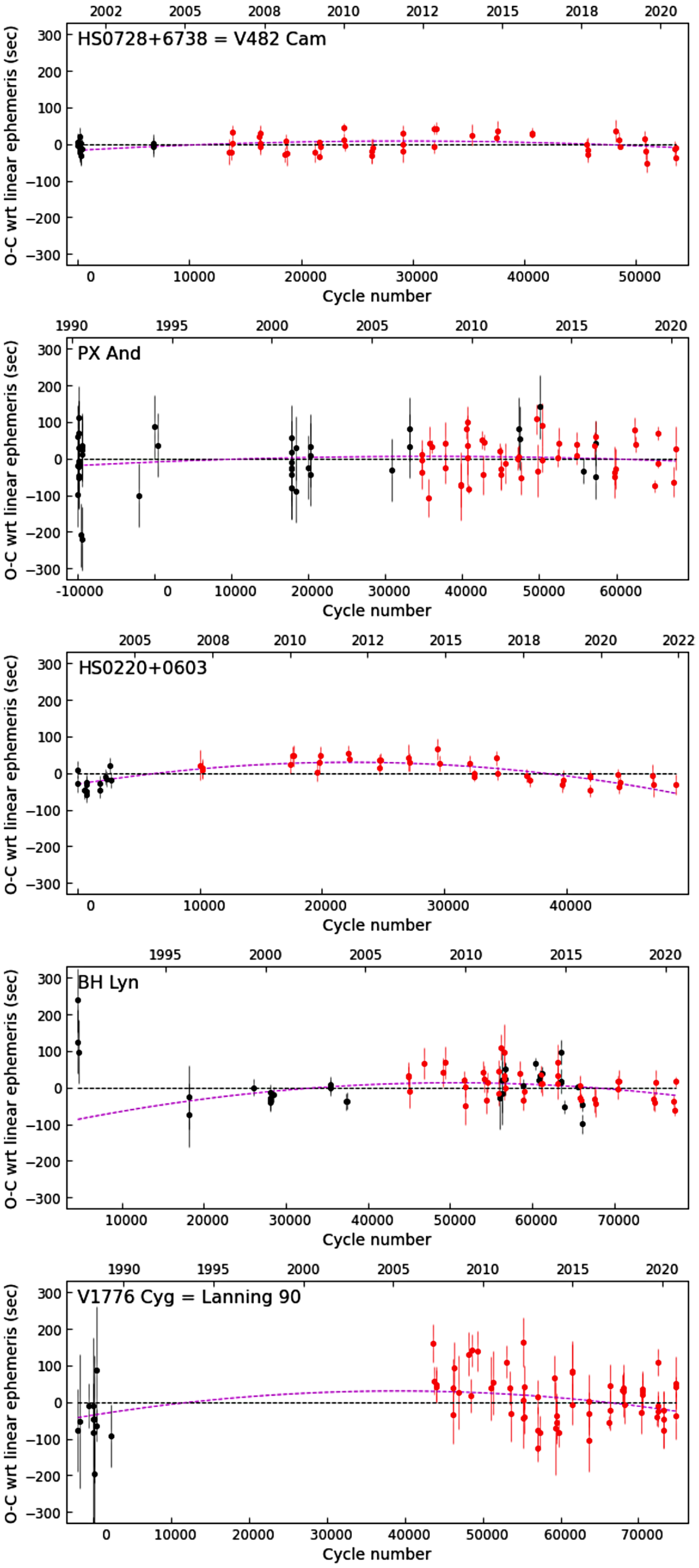

Figure 2. O–C diagrams with linear and quadratic ephemerides for stars showing weak evidence of decreasing orbital period. Data from the published literature or derived from observations in the AAVSO International Database are shown in black while eclipse times measured by the author are shown in red. Linear ephemerides are shown dotted in black, quadratic fits dotted in magenta.

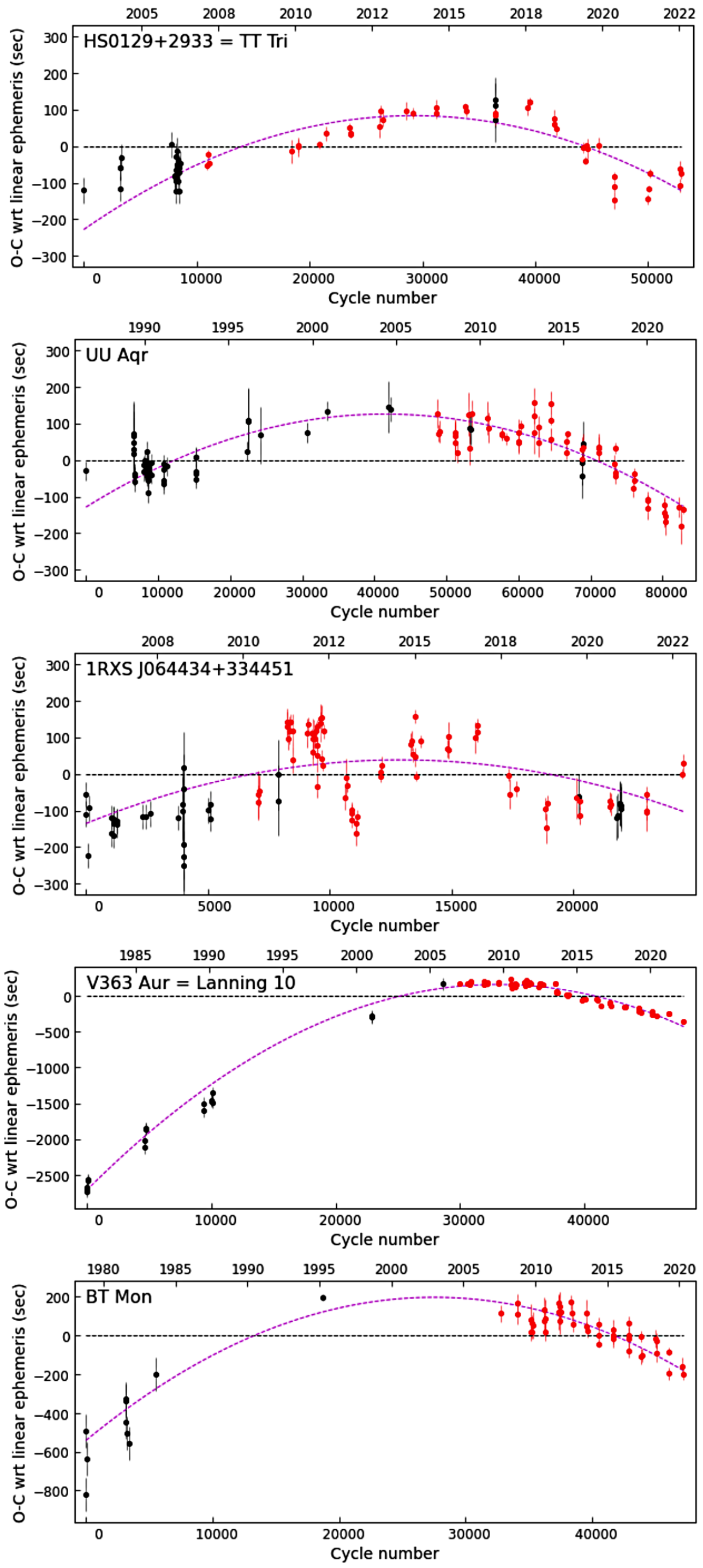

Figure 3. O–C diagrams with linear and quadratic ephemerides for stars showing stronger evidence of decreasing orbital periods. Color coding as in Figure 2.

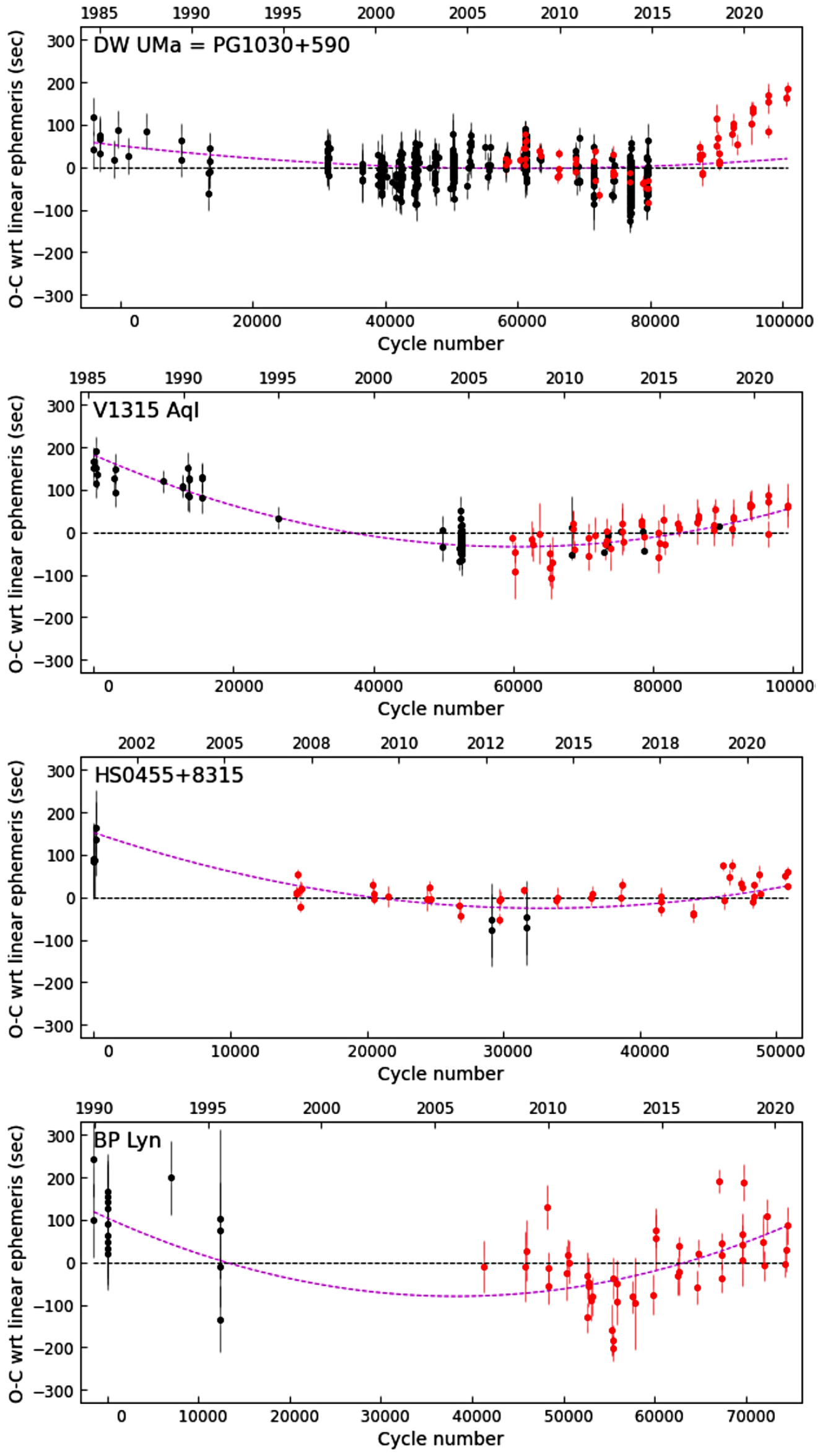

Figure 4. O–C diagrams with linear and quadratic ephemerides for stars showing evidence of increasing orbital periods. Color coding as in Figure 2.

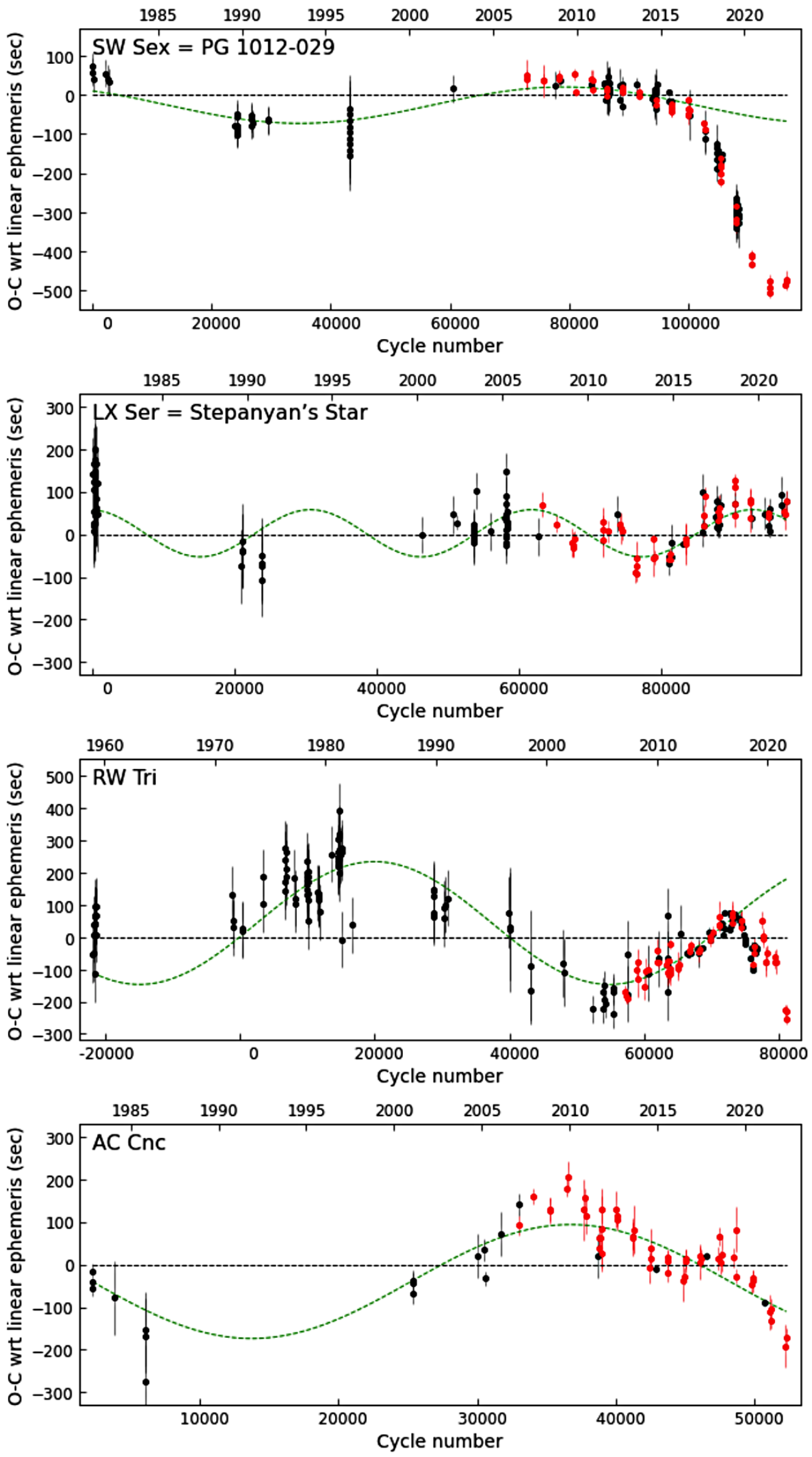

Figure 5. O–C diagrams with linear ephemerides and sinusoidal fits for stars showing evidence of sinusoidal variation in their orbital periods relative to a linear ephemeris. Color coding as in Figure 2 with sinusoidal fits dotted in green.

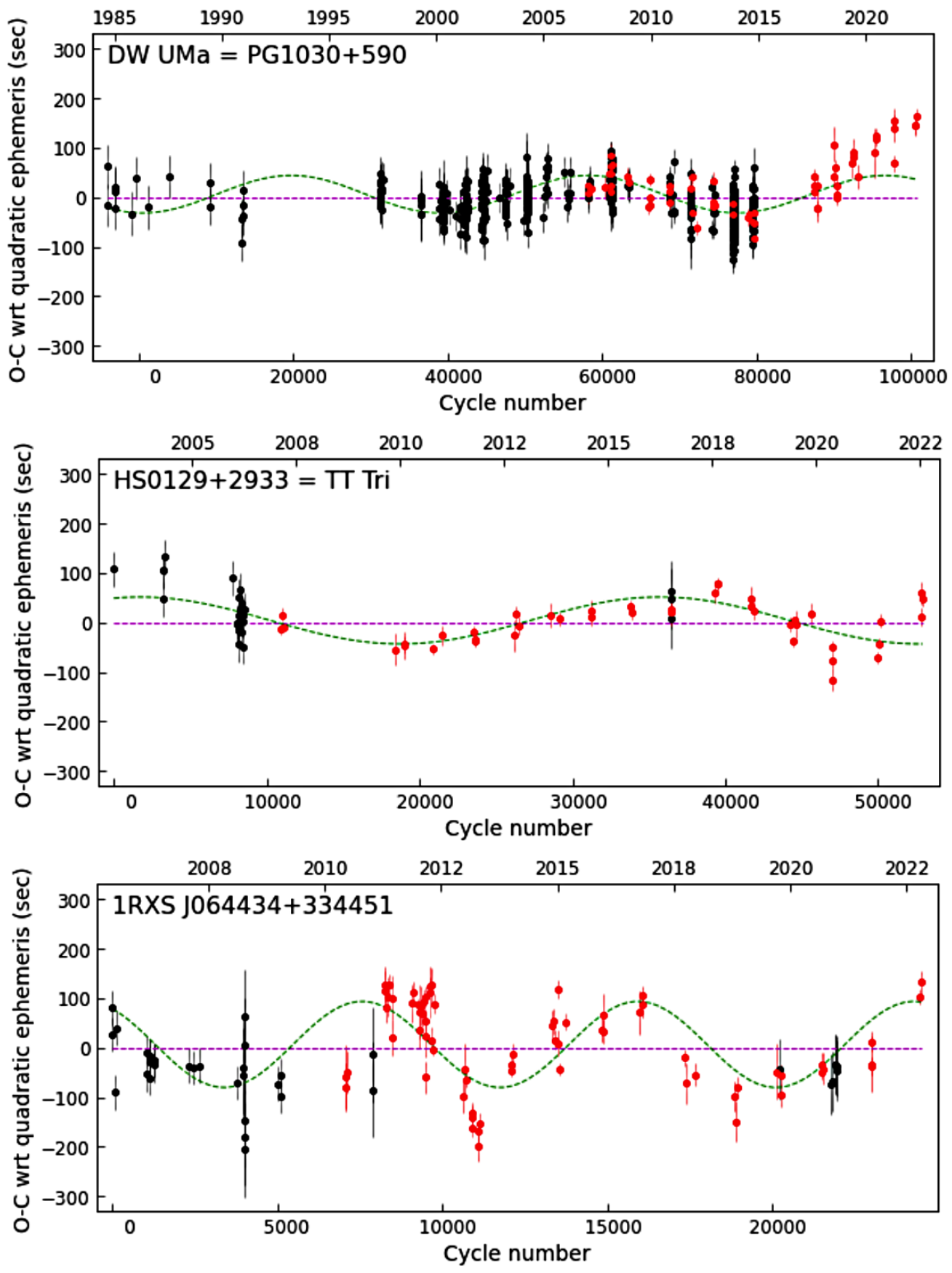

Figure 6. O–C diagrams with quadratic ephemerides and sinusoidal fits for stars showing evidence of sinusoidal variation in their orbital periods relative to a quadratic ephemeris. Color coding as in Figure 2 with sinusoidal fits dotted in green.

## 6. Interpretation

Several mechanisms have been proposed to explain relatively slow changes in the orbital periods of CVs above the period gap, including loss of angular momentum through magnetic braking associated with a magnetized stellar wind (Knigge *et al.* 2011), various versions of the Applegate mechanism associated with magnetically induced changes in the internal structure of the secondary star (Applegate 1992; Völschow *et al.* 2016; Lanza 2020), or a third body in the system whose presence causes a gravitationally induced oscillation of the eclipse time (Qian *et al.* 2013).

We do not believe there has been a sufficiently long period of observations to reach a firm conclusion on the long term behavior of any of the systems reported here. Whether the trends detected so far, as indicated by the fits applied to the O–C data, are maintained in the longer term only further observations will be able to determine. There have been numerous cases in the literature where attempts to assign a specific interpretation to apparently cyclical orbital behavior have failed to stand the test of time (Pulley *et al.* 2022). The dangers of interpreting observations as periodic when only two or three cycles may be present are outlined in Vaughan *et al.* (2016). We therefore do not attempt to assign physical significance to the fits shown in these O–C diagrams, but simply offer our measurements as data to anyone wishing to attempt such an interpretation in the future.

## 7. Summary

We report on a 17-year study to monitor the orbital periods of 18 eclipsing nova-like CVs referred to as SW Sex stars. We added 934 new eclipse times to 1338 times in the published literature and produced an O–C diagram for each star including all available data. This revealed clear trends in the behavior of most of the stars but also that many of the stars experienced deviations from these trends. We observed rapid and unusual decreases of 34 msec (a proportional change of $-2.9 \times 10^{-6}$) in the orbital period of SW Sex during 2017 and of 46 msec (a proportional change of $-2.3 \times 10^{-6}$) in the orbital period of RW Tri during 2018. DW UMa also appears to have recently diverged from the sinusoidal behavior it has been following for the past 30 years. It is clear from these results that observations will have to be maintained over a much longer timescale before definitive statements can be made about their long term behavior, or even whether stable long term behavior is likely for these stars. We intend to continue observing many of these stars.

## 8. Acknowledgements

I am grateful to the anonymous referee for a careful and helpful review. I am also grateful to Boris Gänsicke for his early encouragement to pursue this as a long-term project and to Chris Lloyd for helpful comments on an earlier draft. I acknowledge with thanks the observations of LX Ser by Cook and Dvorak in the AAVSO International Database and the use of AAVSO comparison charts and NASA's Astrophysics Data System. Software developed in the project made extensive use of the Astropy package (Astropy Collaboration 2018).